\documentclass[sn-apa]{sn-jnl}% Math & Physical Sciences, numbered refs

\usepackage{graphicx}
\usepackage{amsmath,amssymb,amsfonts}
\usepackage{mathrsfs}

\usepackage{multirow}
\usepackage{booktabs}
\usepackage{float}
\usepackage{placeins}

\usepackage{algorithm}
\usepackage{algorithmicx}
\usepackage{algpseudocode}

\usepackage{listings}
\usepackage{verbatim}

\usepackage[title]{appendix}
\usepackage{xcolor}
\usepackage{textcomp}
\usepackage{manyfoot}
\usepackage[font=small,skip=0pt]{caption}
\usepackage{bm}

\hypersetup{
  colorlinks=true,
  linkcolor=blue,
  citecolor=blue,
  urlcolor=blue
}

\newcommand*{\bY}{{\bm{Y}}}

\newcommand*{\bR}{{\bm{R}}}

\newcommand*{\bD}{{\bm{D}}}
\newcommand*{\bX}{{\bm{X}}}

\theoremstyle{thmstyleone}

\theoremstyle{thmstyletwo}

\theoremstyle{thmstylethree}

\begin{document}
\title[Synthetic Socio-Economic Index through Autoencoders for Florence]{Developing a Synthetic Socio-Economic Index through Autoencoders: Evidence from Florence's Suburban Areas}
% ---------------- Title & author block ---------------------
%\title[Short Running Title]{Article Title}

\author*[1]{\fnm{Giulio} \sur{Grossi}}\email{giulio.grossi@unifi.it}

\author[1]{\fnm{Emilia} \sur{Rocco}}\email{emilia.rocco@unifi.it}

\affil[1]{\orgdiv{Department of Statistics, Computer Science and Applications “G. Parenti”},
          \orgname{University of Florence},
          \orgaddress{\street{Viale Morgagni 59}, \city{Florence}, \postcode{50134}, \country{Italy}}}


\abstract{
  The interest in summarizing complex and multidimensional phenomena often related to one or more specific sectors (social, economic, environmental, political, etc.) to make them easily understandable even to non-experts is far from waning. A widely adopted approach for this purpose is the use of composite indices, statistical measures that aggregate multiple indicators into a single comprehensive measure. In this paper, we present a novel methodology called AutoSynth, designed to condense potentially extensive datasets into a single synthetic index or a hierarchy of such indices. AutoSynth leverages an Autoencoder, a neural network technique, to represent a matrix of features in a lower-dimensional space. Although this approach is not limited to the creation of a particular composite index and can be applied broadly across various sectors, the motivation behind this work arises from a real-world need. Specifically, we aim to assess the vulnerability of the Italian city of Florence at the suburban level across three dimensions: economic, demographic, and social. To demonstrate the methodology's effectiveness, it is also applied to estimate a vulnerability index using a rich, publicly available dataset on U.S. counties and validated through a simulation study.
}




\maketitle

\section{Introduction}
A ``synthetic index," sometimes also referred to as ``composite", is typically defined in a general sense as a statistical measure that combines multiple variables, often called indicators or elementary (or individual) indices, into a single, unified measure. This general definition highlights that constructing a composite index involves a series of methodological decisions beyond just the aggregation rule. 
Such choices should be grounded in a clear conceptualisation of the phenomenon to be measured and followed by an assessment of the index’s coherence and robustness relative to alternative measures of the same construct, for instance, through comparisons of units’ rankings or other performance metrics, including the stress measure defined in a subsequent section. Consequently, the construction of composite indices remains a subject of considerable interest and ongoing debate in statistics and data analysis. Their widespread use reflects their capacity to summarise complex, multidimensional, and non-directly observable phenomena, while also underscoring the significant methodological challenges they entail.
By aggregating diverse indicators into a single measure, they provide policymakers and the public with an effective tool for understanding and monitoring complex, multidimensional phenomena, as well as for evaluating the effects of policy decisions and interventions within such contexts. Moreover, they enable transparent ranking of the units  to which they refer, such as the geographical areas or other relevant groups of individuals or entities, facilitating comparisons across different spatial and temporal contexts as well as across additional explicitly defined settings. For these reasons, their use spans various fields, ranging from social and economic to environmental and political contexts. They have also been widely adopted by global institutions (e.g. the OECD, World Bank, EU, etc.). The Human Development Index (HDI), the Environmental Performance Index (EPI), the Better Life Index (BLI), the Measure of Economic Well-being (MEW), the Global Competitiveness Index (GCI), the Gender Inequality Index(GII), the European Quality of Government Index(EQGI) are only a few examples of composite indices each tailored to a specific application field. Many other examples are in \cite{bandura2011composite}, which identifies over 400 official composite indices that rank or assess a country according to some economic, political, social, or environmental measures. \\With the growth in availability of data at detailed territorial levels, the computation of composite indices has progressively extended beyond national and regional boundaries. This expansion includes smaller areas, such as municipalities and even sub-municipal zones.\\
Regardless of the specific synthetic index, its scope, or the territorial level at which it is applied, the construction of such an index is based on a data-reduction technique.
There are various examples of data-reduction techniques within the realm of synthetic indices, see \cite{mishra2007comparative, krishnan2010constructing, li2012pca, kotzee2016piloting} among others. Most approaches are based on the use of more or less intricate averages, sometimes weighted and/or even penalized.
The goal of this contribution is to explore the application of autoencoders as a means of dimensionality reduction for datasets comprised of elementary indices. The fundamental aim is to distill essential features from a collection of observations into a singular composite index. The idea of a data-driven composite index is not a novelty. The use of such statistical models can be seen as a nonlinear extension of the construction methods for synthetic indices based on the principal component analysis (PCA). \cite{kramer1991nonlinear} defines the autoencoders as a non-linear version of the PCA. Hence, through the use of autoencoders, our goal is to create a data-driven synthetic index, the ``AutoSynth Index" that, by capturing nonlinear relationships within the data, provides a more accurate representation of the elementary indicators set. \\Although the suggested modus operandi is not tied to the construction of a specific composite index and can be employed in a general manner for constructing a synthetic index in any sector, this work presents a specific case study. It focuses on defining a socio-economic synthetic index to quantify and qualify the multifaceted vulnerabilities embedded within the suburban areas of an Italian municipality, specifically Florence. This index, amalgamating an array of socio-economic and demographic indicators, aims to unveil underlying patterns, identify potential risk factors, and shed light on vulnerability thresholds that may undermine the sustainable development and well-being of suburban communities. It seeks to provide valuable insights for informed policies and interventions to enhance the resilience and prosperity of Florence's suburban areas. \\To demonstrate the AutoSynth Index’s functionality, it was also applied to estimate a vulnerability index for U.S. counties, using the rich dataset provided by \cite{cdcsvi2019}. Additionally, a simulation study was conducted to explore its ability to reproduce the original dimensions within a single feature space across different contexts.\\ The remainder of the paper is structured as follows: Section 2 provides an overview of the preliminaries on the construction of synthetic indices; Section 3 details the proposed methodology; Section 4 presents the motivating case study; Section 5 demonstrates the application of the methodology to estimate a vulnerability index using a large dataset on U.S. counties, while Section 6 explores its performance across various simulated scenarios; finally, Section 7 provides the concluding discussion.
\section{A brief overview of the key steps in constructing a composite index}\label{sec:overview}
Synthetic indices are derived from the aggregation of a set of indicators, each representing a specific dimension of the phenomenon of interest. 
The undoubted advantage of computing such indices, which stems from managing the complexity and multidimensionality of a phenomenon \citep{Mazziotta&Pareto20}, contrasts with what is considered the main limitation of their use. This limitation is the simplification, sometimes deemed excessive, of the object of study, which is argued to inevitably lead to a significant loss of information. Furthermore, while the benefits of using synthetic indices are numerous, so too are the potential errors if the basic and general guidelines that ensure the quality, accuracy, and reliability of the results are ignored. For example, omitting an essential indicator can significantly impact the comprehensive evaluation of the phenomenon of interest. Additionally, the choice of aggregation method is crucial. These are the main considerations that have led some scholars to prefer the dashboards as an alternative analysis method for measuring complex realities. Unlike a synthetic index, a dashboard does not condense the object of study into a single dimension, allowing for the identification of various relevant dimensions. However, it is also clear that this tool lacks the immediate communicative and interpretive capacities that make it easily accessible to users. As noted by \cite{UNECE2019}, composite indicators and dashboards are complementary tools, and the choice between them ultimately depends on the user’s needs.
One way to address the excessive synthesis of a synthetic index and the insufficient synthesis of a dashboard is by using them together. For example, according to \cite{Sachs&all2023} the SDG Index assesses each country's overall performance on the 17 Sustainable Development Goals (SDGs), while the dashboard aids in identifying priorities for further actions and indicates whether countries are on track or off track to achieve the goals and targets by 2030.
An in-depth comparison between the two analysis methods, as well as an exhaustive review of the literature on synthetic indices, will not fall within the scope of this work. However, before describing our approach to constructing a composite index, it is important to briefly outline the main steps to be followed and the key methodological choices to be considered in its development, without claiming to be exhaustive. These steps are described in detail by the \cite{joint2008handbook} and \cite{UNECE2019}, among others.\\
According to \cite{joint2008handbook}, it is first necessary to define the theoretical framework. ``A theoretical framework should be developed to provide the basis for the selection and combination of single indices into a meaningful composite index under a fitness-for-purpose principle." This process should clearly define the phenomenon of interest and its components, while meaningfully involving experts and stakeholders to ensure the relevance and usefulness of the composite index. A fundamental choice concerns the construct's measurement model: whether it should be specified as \emph{reflective} or \emph{formative} \citep{FLEUREN201871, Diamantopoulos}. In a formative specification, the indicators cause and define the latent variable, each capturing a distinct facet of the construct; in a reflective specification, the latent construct gives rise to correlated indicators as manifestations of the same underlying phenomenon. The formative perspective is the most commonly adopted approach in the analysis of socio-economic phenomena and vulnerability \citep{UNECE2019}, where it is typically the socio-economic determinants --- income, demography, social cohesion --- that drive the construct rather than the other way around. This is accordingly the perspective adopted in the present work.
\\
The second step is data selection. The variables or elementary indices should be selected on the basis of their analytical robustness, measurability, coverage for the territorial areas of interest, relevance for the phenomenon to be measured and the relationship between them. 
Data containing large measurement errors or numerous missing values can lead to questionable results. Therefore, the selection of data must be based on a thorough analysis of the data itself. Additionally, various methods for imputing missing data and for handling extreme values should be considered.
Moreover, in addition to carrying out preliminary univariate analysis of the data, it is also necessary to perform a preliminary multivariate analysis to examine the overall structure of the data. This includes checking for correlations and compensability among elementary indices, as well as identifying any redundancy in the information. Compensability refers to the fact that  a unit could compensate for the loss
in one dimension with a gain in another \citep{joint2008handbook,munda2009noncompensatory}.
All these preliminary data investigations are useful for providing insights that guide subsequent methodological choices concerning weighting and aggregation methods.
Normalisation is also usually required before aggregating data, as the indicators in a dataset often have different measurement units. Several normalization methods exist \citep{freudenberg2003composite,jacobs2004measuring}, among which the two most well-known are standardization (or z-scores transformation) and Min-Max normalization. Standardisation converts indicators to a common scale by setting the mean at zero and the standard deviation at one. The Min-Max method normalises indicators to a uniform range of [0, 1] by subtracting the minimum value and dividing by the range of the indicator values.\\
The selection of weights and the aggregation rule are interrelated. 
Weights can generally be considered as coefficients that are attached to individual indices, indicating their relative importance to each other. Their effect on the resulting synthetic index depends on the adopted aggregation method. Most composite indicators rely on equal weighting or the absence of weighting. As outlined in \cite{joint2008handbook} and \cite{greco2019methodological} the two options differ because if the indexes are grouped into a higher order category (e.g., a dimension) and the weights are distributed equally among these dimensions, it does not necessarily imply that the individual indexes within each dimension will receive equal weights. Several other weighting techniques exist. Some
are derived from statistical models, such as principal component analysis or factor analysis, only to mention a few possible methods.  Weights may also be chosen to reflect the statistical quality of the data; for example, lower weighting could be assigned to individual indexes with multiple cases of treated missing data. 
Sometimes the weighting system is subjectively chosen by the developer of the specific synthetic index. To make this choice less subjective, it may involve one or several stakeholders. For a more detailed discussion of the weighting systems, we refer to  \cite{greco2019methodological}\\
The most commonly employed aggregation methods involve various combinations of variables—linear, geometric, or multi-objective—ranging from the simple arithmetic mean to more sophisticated formulas that may incorporate weighted and penalised components. Among these methodologies, the Adjusted Mazziotta–Pareto Index (AMPI) stands out as a non-compensatory composite index designed to measure multidimensional phenomena where indicators are not fully substitutable. Originally developed to assess well-being, AMPI remains a benchmark for evaluating sustainable and equitable well-being (BES) in Italy. Due to its application and inherent properties, it is considered in this work as a potential alternative for comparison with the AutoSynth index proposed herein, which is based on a different aggregation approach utilising a data reduction technique. Such techniques, notably PCA, are employed to construct synthetic indices by reducing the dimensionality of the data while preserving as much variability as possible. For various applications of data reduction techniques in the realm of synthetic indices, refer to studies by \citep{mishra2007comparative, krishnan2010constructing, li2012pca, kotzee2016piloting}, among others.
\subsection{Mazziotta-Pareto Index} In this section, we aim to provide a brief description of the AMPI construction process, referring to \cite{mazziotta2018measuring} for further details. For the construction of the AMPI index, the first step involves normalising the variables under study. This process transforms the non scaled data matrix, $\bX = (x_{ij})$ (where $i=1,\dots,n$  indexes the units and $j=1,\dots,p$  the elementary indices), into a scaled matrix 
 $\bR = (r_{ij})$ using the following formula:
\begin{equation}\label{eq:norm}
    r_{ij}= \left(\frac{x_{ij} - min(x_j)}{max(x_j) - min(x_j)}\right)60 + 70
\end{equation}
In this equation, $min(x_j)$ and $max(x_j)$ represent the minimum and maximum values of the variable $x_j$, serving as the "goalposts" for normalisation. This transformation rescales the original data to a range between 70 and 130, centering the normalized indicators around 100. The choice of these values (60 and 70) is arbitrary but does not affect the ranking of the units and it is already established in the literature by convention.
The values $min(x_j)$ and $max(x_j)$  are referred to as ``goalposts," representing the minimum and maximum reference values. These goalposts can be theoretical (e.g., the unemployment rate cannot exceed $100\%$ or fall below $0\%$) or derived from observed data (e.g., the maximum unemployment rate observed in the sample was $14\%$). 
To incorporate variables that have an ``opposite" polarity relative to the phenomenon of interest, the variable is first normalized as previously described, and then its complement to 200 is calculated.
The third step in constructing the index involves aggregating the normalised variables into a composite indicator as follows:
\begin{equation}\label{eq:ampi}
AMPI_i^{\pm}= \mu_i \pm \sigma_iCV_i 
\end{equation}
where $\mu_i$ is the arithmetic mean of the elementary indicators, $\sigma_i$ is the standard deviation for unit $i$ and  $CV_i$ is the coefficient of variation for unit $i$ .\\
The choice of the operator's sign in equation \ref{eq:ampi}
depends on the nature of the phenomenon being represented. For positive phenomena, such as economic development, the subtraction operator is used. Conversely, for negative phenomena, like social vulnerability, the addition operator is appropriate. This approach penalizes units exhibiting high variability among indicators, ensuring that the index reflects a balanced performance across all considered dimensions.

\section{Constructing synthetic indicators using autoencoders}\label{sec:method}

Our approach aims to develop synthetic indices by employing autoencoders to reduce the dimensionality of datasets comprising numerous elementary indices. 
According to \cite{bank2023autoencoders}, ``an autoencoder is a type of algorithm with the primary purpose of learning an informative representation of the data that
can be used for different applications
by learning to reconstruct a set
of input observations well enough".
\\While various data reduction techniques have been applied in the realm of synthetic indices, such as those by  \citep{mishra2007comparative, krishnan2010constructing, li2012pca, kotzee2016piloting}—these are typically linear methods. Unlike linear techniques like PCA, autoencoders can capture complex, non-linear relationships within the data, allowing for more nuanced feature extraction. Autoencoders, firstly proposed by \cite{rumelhart1986learning}, serve as a nonlinear alternative to PCA, as defined by \cite{kramer1991nonlinear}.

It is worth noting that autoencoders are not the 
only nonlinear extension of PCA available in the literature. A 
well-established family of methods originates from the 
\emph{optimal scaling} framework \citep{gifi1990, deleeuw2009}. 
Categorical Principal Component Analysis (CATPCA, also known as 
PRINCALS) extends classical PCA by simultaneously estimating 
nonlinear transformations of individual variables and the principal 
component solution, using an Alternating Least Squares algorithm 
\citep{linting2012}. Since the transformations are optimised 
jointly with respect to the PCA criterion, CATPCA can effectively 
capture nonlinear relationships between variables --- but only 
insofar as these can be linearised through univariate 
transformations. The resulting model is additive: the latent 
dimension is a linear combination of nonlinearly transformed 
variables, i.e.\ $Z = \sum_j a_j \varphi_j(X_j)$, 
which precludes the representation of interaction effects across 
indicators. CATPCA is particularly suited to mixed-scale data 
(nominal, ordinal, numerical), a setting common in survey-based 
index construction \citep{saukani2019, comim2013human}. A further generalisation 
is OVERALS \citep{vanderburg1988}, which partitions variables 
into predefined sets and maximises homogeneity across them --- a 
structure conceptually akin to the sequential aggregation described 
in Section~\ref{sec:sequential}.

Autoencoders differ in that nonlinearity is 
embedded directly in the reduction model: the encoder maps the 
entire input vector through compositions of activation functions 
across hidden layers, yielding a latent representation 
$Z = f(X_1, \ldots, X_p)$ that can capture arbitrary nonlinear 
interactions among indicators, not only additive effects of 
individual transformations. In this sense, autoencoders generalise 
the optimal scaling approach rather than merely offering an 
alternative to it.

Autoencoders are neural network architectures designed to learn efficient representations of data by reconstructing the original input matrix $\bX$, while constraining the encoding to a lower-dimensional subspace. The input matrix $\bX$ has dimensions $N\times p$, where $N$ represents the number of observations and $p$ denotes the number of features (or elementary indicators) in the dataset. The objective is to extract a compressed representation that retains the most relevant information. The target vector $\widetilde{\bY}$, a $N \times 1$ vector, represents the essential output or label associated with the observations. Autoencoders are particularly useful for identifying and learning the most significant features from high-dimensional data, making them powerful tools for dimensionality reduction and unsupervised learning in the context of neural networks. An autoencoder comprises two primary components: an encoder and a decoder. The encoder function 
$\phi$ compresses the input data $\bX$  into the lower-dimensional latent representation  $\widetilde{\bY}$, such that $\widetilde{\bY} = \phi(\bX)$, capturing the most salient features. Subsequently, the decoder function $\psi$ endeavours to reconstruct the original input from this compressed form, yielding $\widetilde{\bX} = \psi(\widetilde{\bY})$. 
 Consequently, the overall transformation is represented as $\widetilde{\bX} = \psi(\phi(\bX))$.
%\begin{equation}\label{eq:encoderdecoder}
    \begin{align}\label{eq:encoderdecoder}
\text{Encoder:} \quad & \widetilde{\bY} = \phi(\bX) = \sigma(W\bX + b),\nonumber \\
\text{Decoder:} \quad & \widetilde{\bX} = \psi(\widetilde{\bY}) = \sigma'(W'\widetilde{\bY} + b'),
\end{align}
%\end{equation}
where $W$ is a set of activation weights, $b$ is a bias vector and $\sigma$ is a proper activation function. This process enables the model to learn efficient codings of the data in an unsupervised manner, making autoencoders particularly suitable for dimensionality reduction tasks. For a more detailed description, refer to \cite{lecun2015deep}. 
We provide additional explanations for our modelling choices in section \ref{sec:inw}.

% \textcolor{red}{Forse occorre dire quacosa su $W$ e $b$ come li assumiamo noi e relazione con pesi input e output ...}\commentG{Ho rimandato alla sezione dove si parla di questo aspetto}

\begin{figure}
    \centering
    \includegraphics[width=.7\textwidth]{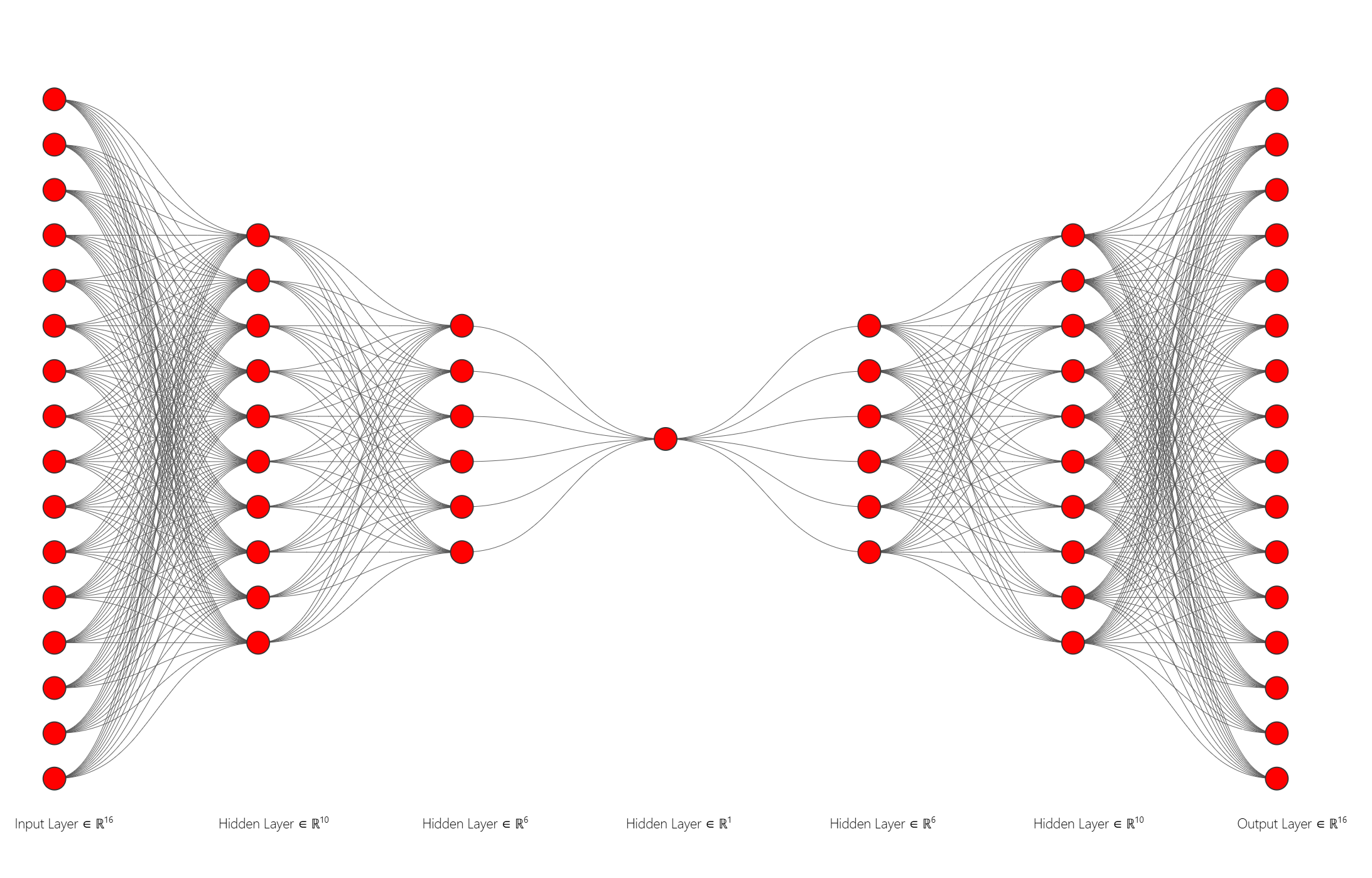}
    \caption{Basic scheme of autoencoders. In this application, inputs will be  indicators of socio-economic development, while the code will be the synthetic indicator}
    \label{fig:diagram}
\end{figure}

Central to the autoencoder architecture is the optimisation of a specific objective function. This function aims to minimise the discrepancy between the original input matrix $\bX$ and the reconstructed output $\widetilde{
\bX}$. The distance metric $D(\bX, \widetilde{\bX})$ quantifies this discrepancy. The optimisation target is formally articulated as:
\begin{equation}\label{eq:argmin}
\underset{\phi, \psi}{\operatorname{argmin}} | \bX - \psi(\phi(\bX))|_\bD
\end{equation}
See figure \ref{fig:diagram} for a graphical representation of the autoencoder architecture.

% To achieve data compression, we can leverage the estimated encoder model, denoted as $\widehat{\phi}$, to transform the input data,
% taking an input matrix and maps it into a single vector $\bY$ of dimensions $N \times 1$. The transformation is expressed as:

% $$\widehat{\bY}_{N \times 1} = \widehat{\phi}(\bX_{N \times p})$$

The choice of the metric distance $D(\bX, \widetilde{\bX})$ can affect the results and thus its choice should be handled with care. In our work, we refer to the term distance as Euclidean distance, which is a baseline choice. Other distance measures are possible, think for instance to fuzzy distance measures, but we leave the discussion of composite index based on different metrics to future work. 
Additionally, it is necessary to choose a proper activation function for the nodes of the neural network. To grasp the nonlinearities that could be present in the dataset, we opt for a Rectified Linear (ReLu) activation function throughout this work.
$$ \operatorname{ReLu}(x_i) = \operatorname{max}(0,x_i)$$

As underlined by \cite{michelucci2022introduction}, ReLu function is particularly suitable when the values in the dataset $\bX$ are in the very majority positive, this is usually the case of indicators in social research, that can be normalized but usually in a positive range, see for instance the normalization proposed by \cite{mazziotta2018measuring}, that spans between 70 and 130. 
We tested other nonlinear options (logistic activation, hyperbolic tangent), but we find the most promising results with ReLu.  
For an in-depth discussion on the choice of activation functions in autoencoders, see also \cite{klopries2023flexible}.

Five additional considerations regarding the use of autoencoders for constructing synthetic indices include: the potential incorporation of input weights; the interpretation of Indicator relevance; the hierarchical organization of more synthetic indices; the issue of the index's polarity not definable a priori and finally, defining a criterion to assess AutoSynth's performance relative to other aggregation methods
\subsection{Input weights}\label{sec:inw}
Researchers might occasionally prefer to define a set of initial weights proactively, rather than allowing the autoencoder to determine weights from the data. This approach is particularly relevant when aiming to accurately depict phenomena where specific elementary indices require distinct weighting relative to others. Such scenarios align with the principles of synthetic index construction utilizing weighted averages, where expert knowledge influences the initial weight assignment.
Notice that in equation \ref{eq:encoderdecoder}, a set of weights is present. The vector $W$ represents the coefficients applied to the consecutive combination of the set of elementary indices $\bX$ through the layers. 
In practice, we can emphasize the importance of some elementary indices in the reconstruction of the output by specifying a set of input weights that weigh in an asymmetric way the indicators used in the autoencoder. 
This specification is very important when the researcher has the availability of prior information about the relative importance of the indicators used to construct the synthetic index.
In particular, this allows us to derive an informed representation of the latent phenomena, which is different from the representation that we could get if we have no prior information about the importance of elementary indices. 
On the other hand, a specification of the set of weights that is too unbalanced, or narrow, associated with an autoencoder architecture not flexible enough, could lead towards representations of the latent phenomena that are biased or imprecise. 
It is worth noticing that the use of input weights resembles the use of weights in the construction of synthetic indices with the common usage methods, in which the expert has control (and responsibility) over the set of weights and, intrinsically, over the output of the analysis. 
From this perspective, this procedure within the Autosynth represents an advancement, as we incorporate the expert knowledge with the nonlinear data compression through the autoencoder. 
In this work, we remain agnostic with respect to the input weights, setting all of them equal to one, and with respect to the bias vector $b$, setting it equal to zero. However, different specifications are possible, but it is not our primary focus to treat them here, as their specification are case specific.

\subsection{Post-estimation indicators relevance}
On a different perspective, it is possible to extract the \textit{posterior} importance of an elementary index, namely, how much the indicator is affecting the synthetic index. 
Here, we propose to estimate these Indicator relevance by estimating the reconstruction error associated with each indicator employed.  By examining these errors, we assess each indicator's relative weight and potential disproportionate influence due to correlation structures, refining the indicator set for the synthetic index construction.
The process of calculating the reconstruction error is as follows:

\begin{equation}\label{eq:posterior}
\begin{aligned}
    \varepsilon_p = \frac{1}{N}\sum_{i=1}^N |\bX_{ip} - \widetilde{\bX}_{ip}| \\
\zeta_p = \frac{\varepsilon_p}{\sum_{p=1}^K \varepsilon_p}
\end{aligned}
\end{equation}

Where $\widetilde{\bX}$ is the indicator matrix reconstructed through the encoder and decoder function, as in equation \ref{eq:encoderdecoder}. 

\subsection{Sequential autoencoders}\label{sec:sequential}
The utilisation of synthetic indices is particularly vital for aggregating elementary indices across various domains. In such cases, a bifurcated approach to aggregation is highly effective. Initially, variables within identical domains are merged—for example, combining all indicators related to economic vulnerability into a single economic vulnerability index. Following this, the second phase involves the consolidation of vulnerability indices from distinct domains.
This idea is equally applicable to the Autosynth tool, where we could institute a hierarchical aggregation process. Initially, variables within the same domain are aggregated, and subsequently, variables across different domains are merged. This hierarchical structure facilitates the derivation of intermediate-level indices, allowing for an in-depth analysis not just at the synthetic index level, which represents the overall issue, but also within more separate domains.
Moreover, it's crucial to acknowledge that computational costs can escalate with a significant volume of observations, particularly when numerous elementary indices are involved. Therefore, segmenting the aggregation process into two distinct phases not only offers methodological benefits but also enhances calculus's efficiency.
This hierarchical aggregation strategy bears a structural resemblance to the set-based approach of OVERALS \citep{vanderburg1988}, where variables are likewise grouped into substantive domains and aggregated in stages. The key difference is that OVERALS is constrained to maximise linear canonical correlations between sets of transformed variables, whereas sequential autoencoders can capture nonlinear cross-domain interactions through the composition of encoding functions.

\subsection{Polarity}
Since autoencoders project the encoded representation of the data matrix $\bX$ into a latent subspace, the resulting values may differ in scale and even in ordering compared to the original data. As a consequence, the direction of the resulting index is not necessarily preserved. This situation can compromise the interpretability of the index, as its polarity is not defined a priori \citep{mazziotta2013methods}, namely whether higher values represent a desirable or undesirable phenomenon. Therefore, the results obtained through aggregation with Autosynth, similarly to those from PCA) are not directly interpretable. 
To address this issue, the researcher can rely on domain expertise to assess whether the polarity of the latent dimension is consistent with expectations or alternatively, use an external synthetic index as a reference for interpretation. In this work, we adopt the synthetic index derived from the AMPI methodology as a \textit{compass}: if the unit with the highest AMPI score does not fall within the first quartile of the Autosynth-based index, we invert the polarity of the Autosynth index.   

\subsection{Measuring Autosynth performances}\label{sec:stress}
To evaluate Autosynth’s performance, we assess two criteria: (i) the consistency of its results with those obtained from alternative indices—particularly in terms of unit rankings—and (ii) a stress measure, which quantifies how well the low-dimensional representation preserves the original distances between observations in the original dimension, and consequently how accurately it reflects the relationships among units.

For the first evaluation criterion, we selected two benchmark indices: the AMPI and a PCA-based index. AMPI is a non-compensatory synthetic measure constructed by aggregating elementary indicators and is widely used in Italy to summarise multivariate phenomena similar to our motivating case study. The PCA-based index was chosen because, like AutoSynth, it reduces data dimensionality while retaining maximal variance, thus serving as the linear counterpart to our proposed method.

Concerning the second criterion, we employ the two-dimensional stress measures introduced by \cite{kruskal1964nonmetric}, as represented in Equation \ref{eq:stress}.

\begin{equation}\label{eq:stress}
 \Theta = \sqrt{\frac{\sum_{i=1}^N (d_{ij} - \widetilde{d_{ij}})^2}{\sum_{i=1}^N d_{ij}^2}} \end{equation}

Here, $d_{ij}$ represents the Euclidean distances between units $i$ and $j$ in the matrix of elementary indicators $\bX$ as $d_{ij} = \sqrt{\sum_{k=1}^{p} (x_{ik} - x_{jk})^2}
$, while $\widetilde{d}_{ij}$ denotes the Euclidean distances between the same units in the synthetic index $\widetilde{\bY}$ as $\widetilde{d_{ij}} = \sqrt{(\widetilde{y}_{i} - \widetilde{y}_{j})^2}
$.
By construction, the stress measure will lie between 0 and 1, and lower values for the index represent a better representation of the original outcome.

\section{Assessing vulnerability in Florence}\label{sec:florence}
In this section, we present the case study that inspired our research: the development of a Socio-Economic and Demographic vulnerability Index (SEDI) specifically designed for the Florentine suburbs.

Cities across the globe are undergoing rapid transformations driven by urbanization and societal shifts. These dynamics create a complex interplay between social, economic, and demographic factors, posing significant challenges for researchers, policymakers, and urban planners. 
The historic city of Florence, Italy, provides a compelling case study for examining these global trends. Despite its world-renowned cultural heritage, Florence is not immune to the challenges of urban evolution. The city's evolving social fabric has led to significant disparities in residents' living conditions, particularly within its suburban areas, raising the need for careful analysis and strategic solutions to ensure the well-being of all Florentine residents.
A powerful approach to shed light on these complexities lies in the construction and evaluation of a Socio-Economic and Demographic vulnerability Index (SEDI) at the suburban level. This study proposes the development of such an index specifically tailored to the Florentine suburbs.  The SEDI will integrate a range of socio-economic and demographic indicators to quantify and qualify the multifaceted vulnerabilities within these communities. This will be achieved by employing the proposed autoencoder-based aggregation method described in the previous section.
This analysis aims to identify risk factors that may hinder sustainable growth and diminish the quality of life for residents. The SEDI will define critical vulnerability thresholds, supporting policymakers with essential data to design targeted interventions. By highlighting the specific vulnerabilities of Florence's suburban areas, the SEDI will guide the creation of tailored policies and measures aimed at strengthening social cohesion, enhancing economic opportunities, and improving overall community well-being. In doing so, the proposal will lay the foundation for more sustainable and equitable urban development in Florence.

\subsection{Three pillars for vulnerability} \label{sec:pillars}

Urban vulnerability has been operationalised along three main lineages \citep{romerolankao2011}: a \emph{natural hazards} approach, centred on exposure to environmental threats; an \emph{inherent vulnerability} perspective, emphasising the socioeconomic, demographic, and institutional conditions that constrain coping capacity; and a \emph{resilience} perspective, framing cities as coupled human--environment systems. Each lineage implies a different domain structure. The CDC/ATSDR Social Vulnerability Index \citep{flanagan2011}, for instance, groups 16 census variables into four themes --- socioeconomic status, household composition and disability, minority status, and housing type and transportation --- to support disaster preparedness, while broader frameworks such as the Social--Ecological--Technological Systems approach \citep{mcphearson2016} further integrate ecological and infrastructural dimensions.

The SEDI index proposed in this work is positioned within 
the inherent vulnerability tradition, with a specific focus 
on socioeconomic and demographic conditions at the 
sub-municipal scale. The choice of three domains --- 
economic, demographic, and social --- follows the established 
framework for sub-municipal vulnerability assessment in 
Italy, adopted in comparable studies for Bologna 
\citep{comunebo}, Naples \citep{comunena}, and other 
Italian municipalities \citep{davino2021measuring, tronu2020istat, busetta2010socio}
. This tradition is oriented towards informing 
local social policy and urban planning rather than assessing 
exposure to environmental hazards.

One of the primary objectives of public policy is to address vulnerabilities within the population. Developing tools to support this goal is an evolving focus within the fields of social statistics and public policy, as highlighted by  \cite{saisana2012sustainable} and \cite{khan1991measurement}. 
In recent years, a substantial body of research has emerged, focusing on the measurement of these intricate concepts, resulting in the development of a wide range of synthetic indicators.
Following previous works on the socio-economic and demographic vulnerability in Italy \citep{tronu2020istat, busetta2010socio}, even at sub-municipal level, both in Italy \citep{comunebo, comunena, davino2021measuring}, and in Europe \citep{von2023index, martinez2024remaking}, we adopt a theoretical framework based on three sub-pillars for SEDI: economic vulnerability, demographic vulnerability and social vulnerability.

The study of demographics in evaluating a territory's vulnerability is grounded in assessing the population's needs within its social context. Specifically, we can identify at least three major demographic factors that can be interpreted as precursors of vulnerability for a social environment: ageing population, low birth rate and depopulation. 
The gradual ageing
of the population is a well-known phenomenon that is transforming the social and economic landscape of the 21st century \citep{christensen2009ageing}, placing increasing pressure on policymakers to enhance healthcare services and improve the living conditions of older adults \cite{skouby2014smart}. The evolution of Italian demographics, in particular, suggests that this is a central theme for the planning and management of a territory \citep{reynaud2019population}. Consequently, identifying areas of the city most at risk due to population aging is essential for designing targeted interventions and ensuring the provision of facilities that adequately respond to this demographic challenge.
The second dimension representing a demographic challenge is the low birth rate, as highlighted by \cite{billari2004patterns}. Previous analyses conducted in other metropolitan areas and several ISTAT reports have highlighted the vulnerability associated with the imbalance between births and deaths, with particular emphasis on low birth rates. Consequently, we adopt the ``natural balance" as an indicator to capture this specific demographic dimension. 
The third factor indicative of demographic vulnerability is the depopulation of certain areas. This phenomenon becomes particularly evident in studies involving comparisons between municipal areas, where disparities between urban centers and inland regions highlight the trend towards depopulation \citep{pinilla2008rural, vendemmia2021institutional}. 
At the sub-municipal level, the effects of depopulation are likely less pronounced, but nonetheless significant.  The deterioration of the social cohesion of a neighborhood can push people to move away from it and to relocate to more attractive residential areas, consequently, the decrease in population in an area not due to the natural balance can be interpreted as a loss of attractiveness of the area and regarded as a sign of vulnerability.

Focusing on the economic aspects, we identify the relative poverty 
indicator and the indicator of insufficient capital accumulation, 
proxied by the share of citizens paying rent, as major sources of 
vulnerability. The relationship between poverty and vulnerability is 
well documented in the literature \citep{adger2014vulnerability, coulthard2018multiple, staveren2014last}, with higher vulnerability typically observed in 
economically disadvantaged areas. Beyond income-based measures, we 
include the share of renters as a proxy for long-term asset 
deprivation. In the Italian institutional context, where the 
homeownership rate exceeds 74\%, real-estate 
property constitutes the primary vehicle for household wealth 
accumulation and intergenerational transfer 
\citep{ferrante2021, guiso2002}. At the area level, a high share 
of renters therefore signals lower aggregate capital accumulation, 
capturing a complementary dimension of economic vulnerability that 
current income alone does not reflect. We note that direct measures 
of household wealth are not available at the sub-municipal level in 
Italy, making housing tenure the best available proxy at this 
territorial granularity.
Additionally, we include median individual and median family income, as lower income levels are consistently associated with reduced coping capacity and heightened vulnerability at the area level \citep{cutter2003social, davino2021measuring}.

\begin{comment}
Focusing on the economic aspects, we identify the relative poverty indicator and the indicator of insufficient capital accumulation, proxied by the share of citizens paying rent, as major sources of vulnerability. 
The relationship between poverty and vulnerability is well documented in the literature \citep{adger2014vulnerability, coulthard2018multiple, staveren2014last}, with higher vulnerability typically observed in economically disadvantaged areas.. Beyond this established relationship, we also consider the insufficient accumulation of capital required to purchase a home as an additional, significant, and often overlooked factor contributing to economic uncertainty and vulnerability.
\end{comment}

To capture social vulnerability, we employ a set of variables, among which the presence of elderly residents living alone emerged as a key indicator, given their often greater need for health and social care \cite{victor2000being, golden2009loneliness, roh2022living}. Additionally, we consider the vulnerability of minors in single-parent households, who may require greater social protection and assistance \citep{bianchi2010work, bianchi2014changing}. Minors from foreign-origin families are also included, given the potential challenges they face in integrating into Florentine schools and the broader social fabric \citep{scardigno2019cultural, bloemraad2023unpacking}. Moreover, drawing on prior research \citep{davino2021measuring}, we assume that higher levels of educational attainment are associated with lower social vulnerability, insofar as they enhance resilience at both the individual and community levels. Accordingly, we include the percentage of graduates residing in each area as an indicator. Finally, the proportion of vacant housing units is incorporated as a proxy for potential neighborhood abandonment, a condition frequently linked to increased social vulnerability.

The interplay across these dimensions is documented in the literature: for instance, it has been observed premature aging across lower income classes in \cite{steptoe2020lower}, or correlations across depressed areas with sizable integration issues \citep{majid2020global, chakraborty2020place}. 

A comprehensive vulnerability framework would ideally encompass additional dimensions beyond the three considered here. A \emph{health} dimension --- hospitalisation rates, chronic disease prevalence, access to healthcare --- is well established in the social determinants literature \citep{flanagan2011, cutter2003social, khazanchi2020county}. An \emph{ecological} dimension --- urban heat islands, flood risk, air quality, green space --- is central to the environmental vulnerability tradition \citep{romerolankao2011}. A \emph{cultural} dimension --- access to cultural services, community participation --- would further capture social cohesion and collective resilience. However, none of these indicators are currently available at the sub-municipal level in Italy from the administrative sources employed here, and the three domains retained for the SEDI represent the best feasible operationalization at the required territorial granularity.

\subsection {The data}
Almost all the indicators mentioned in Section \ref{sec:pillars} are based on data collected during 2021. Demographic and social indicators are sourced from the Civil Registry of Florence while economic indicators are provided by the Italian Revenue Agency (Agenzia delle Entrate - AdE) and further elaborated by the Municipality of Florence. The only indicators not referring to the year 2021 are the percentage of graduates and the percentage of unused dwellings that are derived from the 2011 census. 
Table \ref{tab:my_label} presents the main descriptive statistics related to all elementary indices considered along with their respective sources.
\begin{table}[h]
    \centering
      \resizebox{\textwidth}{!}{ 
\begin{tabular}{ccrrrrrrrr}
\toprule
\textbf{Domain}  & \textbf{Elementary Index}  & \textbf{Mean} & \textbf{st.dev.} & Min & 25\% & 50\% & 75\% & Max & Source\\
\midrule
Demography & \% Over 80 & 9.600 & 2.085 & 1.261 & 8.420 & 9.680 & 10.964 & 15.028 & Fl. civil registry\\
& $\Delta$ population & $-$2.867 & 3.471 & $-$11.337 & $-$4.430 & $-$2.942 & $-$1.563 & 16.098 & Fl. civil registry\\
& Natural Balance & $-$29.310 & 21.070 & $-$94.200 & $-$40.900 & $-$26.600 & $-$14.850 & 4.000 & Fl. civil registry\\
\hline
Social & \% Over65 living alone & 9.210 & 1.733 & 0.840 & 8.262 & 9.500 & 10.083 & 12.973 & Fl. civil registry\\
& \% Under18 foreigners & 16.279 & 8.277 & 2.308 & 10.586 & 14.529 & 19.657 & 39.159 & Fl. civil registry\\
& \% Under 18 -- Single parent & 42.580 & 5.720 & 18.520 & 39.110 & 42.180 & 44.920 & 64.780 & Fl. civil registry\\
& \% Unused dwellings & 3.578 & 2.737 & 0.271 & 1.684 & 2.976 & 4.647 & 14.506 & 2011 Census\\
& \% Graduated & 37.490 & 10.445 & 14.800 & 30.460 & 38.170 & 46.330 & 57.080 & 2011 Census\\
& \% Pop. circulation  & 1.613 & 2.099 & 2.917 & 3.275 & 3.585 &  3.928 & 15.918 & Fl. civil registry\\
\hline
Economic & \% people under poverty line & 32.970 & 3.751 & 21.170 & 30.660 & 32.700 & 34.710 & 42.880 & AdE\\
& \% rents & 20.628 & 6.721 & 8.092 & 15.711 & 19.269 & 25.226 & 37.841 & 2011 Census\\
& Median Income (individual) & 20801 & 2207 & 16164 & 19340 & 20564 & 21622 & 29048 & AdE\\
& Median Income (family) & 28979 & 4255 & 19984 & 26280 & 28268 & 30919 & 47774 & AdE\\
\bottomrule
\end{tabular}
}
    \caption{Descriptive statistics of the elementary indexes -- Mean, standard deviation, minimum, maximum, median, first and third quantile.}
    \label{tab:my_label}
\end{table}

As units of observation, we assume the $N=74$ suburban units into which the area of Florence is partitioned. 
These units represent a middle-level aggregation between the census areas and the broader administrative districts of Florence, which would be too large for the scope of this study. Even if these units stem from administrative sources, they represent homogeneous partitions of the city, particularly relevant for our purposes. 
Two of them were excluded from the analysis, as their population is too scarce to have reliable estimates (under 100 inhabitants).

% \textcolor{red}{Forse occorre dire quacosa su normalizzazione e calcolo dell'indice...} \commentG{si ma nel paragrafo dopo}

\subsection{Results}

In this section, we present the estimation and discussion of SEDI index using the Autosynth methodology for the suburban Florentine case study. 
It is worth noting that we first scale the original outcome using the formula described in equation \ref{eq:norm}, and then we applied the Autosynth to this dataset, by specifying equal weights ${W}$ and bias vector $b=0$.

%\textcolor{red}{riportare che abbiamo fatto diverse run e riportiamo la mediana della distribuzione degli indici sintetici}
 \begin{figure}
     \centering
     \includegraphics[width=1\textwidth]{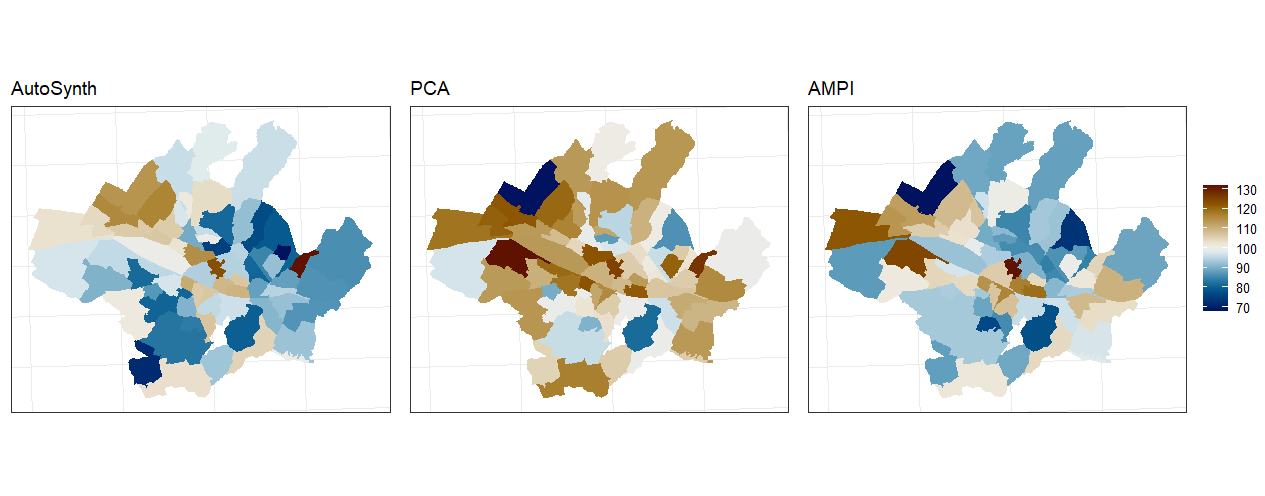}
     \vspace{-0.3cm}
     \caption{AMPI, PCA and AutoSynth vulnerability Index for Florence, normed data}
    % \label{fig:id_index_fi_norm}

          \centering
     \includegraphics[width=1\textwidth]{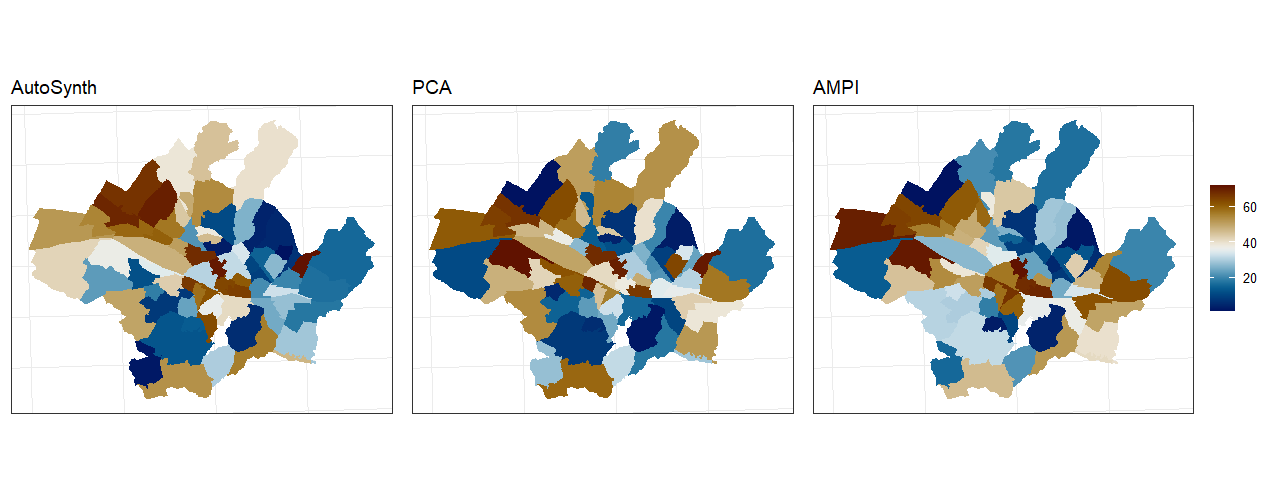}
     \vspace{-0.3cm}
     \caption{Rank statistics for the AMPI, PCA and AutoSynth vulnerability Index for Florence, normed data}
     \label{fig:index_fi}

 \end{figure}

The analysis of the results presented in Figure \ref{fig:index_fi}, allows us to evaluate the performance of the SEDI in capturing the latent vulnerability across the sub-municipal areas of Florence and to compare it with the two reference indices: AMPI and the PCA-based index. As specified in the previous sections, while PCA and AMPI do not allow for an uncertainty evaluation, Autosynth is an inherently stochastic methodology and thus, we obtain a distribution of results by iterating 500 times the calculation. Here we report the median of the distribution of results. Discussion of the results is based on the SEDI index we got from Autosynth, as PCA and AMPI are reported only for comparative purposes. 

%\textcolor{red}{specificare che nei risultati si commenta i risultati con autosynth}

A visual examination of the results reveals that the areas exhibiting the greatest vulnerability are the historic city center and the western parts of Florence, see table \ref{tab:top5_florence}. 
On the other hand, the least vulnerable areas are located in the eastern suburbs and in the south area, which is a worldly known area for the beauty of the landscape, as depicted in table \ref{tab:top5_florence}. 
These findings are consistent with previous analyses conducted on the same case study by \cite{dugheri2023assessing}.

It may seem counterintuitive that one of the areas facing the most significant socio-economic and demographic challenges is the historic center, a UNESCO World Heritage Site known worldwide. However, one must consider the gentrification that has occurred in this area: alongside historic residences, there are older and more affordable apartments, often inhabited by students or immigrants who cannot afford renovated housing. Additionally, the historic center has, in recent years, experienced widespread conversion of many apartments into Airbnbs rentals, reducing the number of dwellings available to permanent residents.

Another area that emerges as more fragile is West Florence. This area has exhibited systemic vulnerabilities for years, being on average one of the areas with the lowest median income in the city. However, unlike the historic center—which is characterized by significant economic disparities—West Florence presents a more uniform profile, with widespread challenges but fewer instances of social marginalization.

By comparing the results from the three methods, we observe that all the methods produce reasonably similar estimates. In particular, Autosynth and PCA produce very similar estimates, leading to similar conclusions over composite vulnerability in Florence. This results is confirmed both in the absolute values of the index, which are similar, but most and more importantly in the ranks across the sub-municipal areas, which are a crucial point in the analysis of the composite indicators. 

\begin{figure}
    \centering
    \includegraphics[width=.5\linewidth]{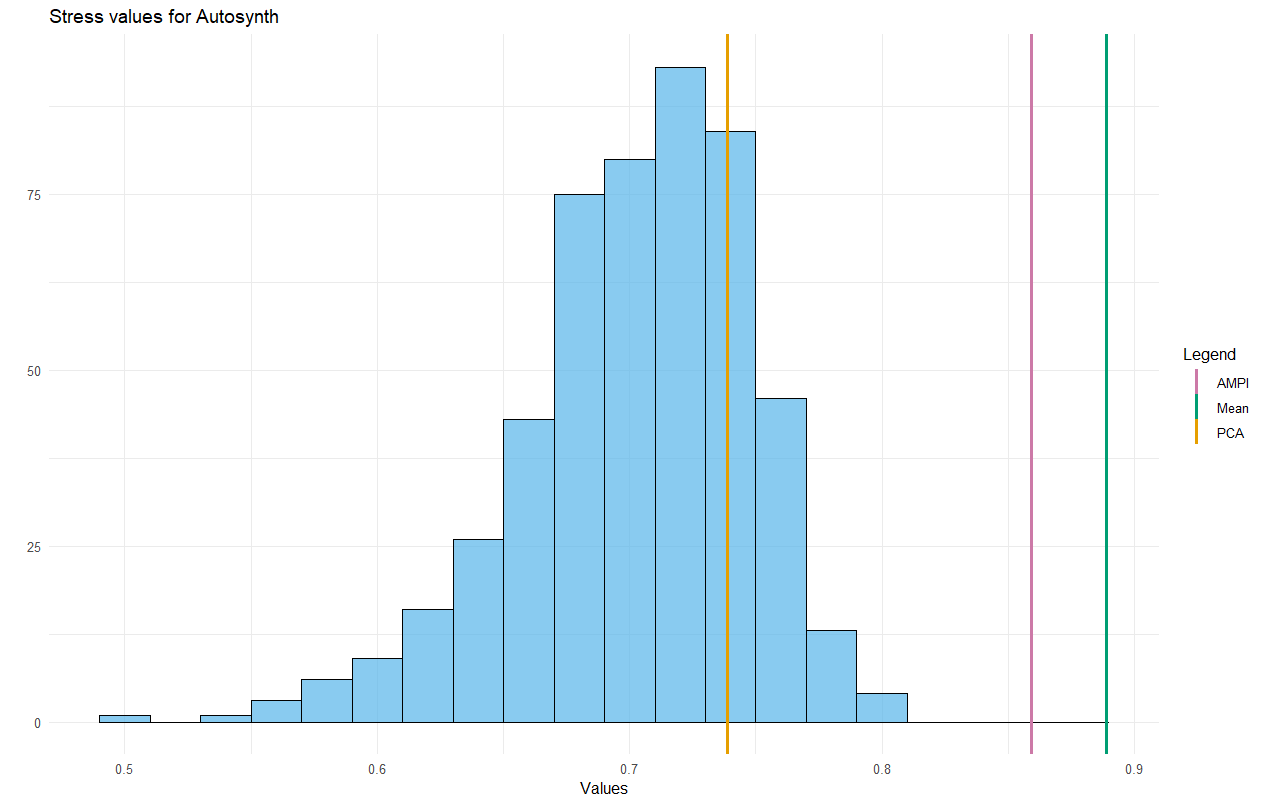}
    \caption{Stress values for autosynth index, compared to the other methods - normed data}
    \label{fig:stress_norm_florence}
\end{figure}

This is not unexpected, as both techniques aim to reduce information from a multidimensional space to a one-dimensional one. In contrast, the results obtained through the AMPI differ noticeably in the absolute values of the vulnerability index. Despite this discrepancy, the ranking of the sub-municipal areas remains quite similar across all three methods. 
Figure \ref{fig:stress_norm_florence} shows that the majority of autosynth samples have a lower stress with respect to the alternative methods. We conclude that the ordering provided by the SEDI index is not particularly sensitive to the chosen aggregation method, but that AutoSynth offers better performance in reproducing the original information.

\begin{table}[]
    \centering
\begin{tabular}{lrr}
\toprule
\textbf{Elementary Index}  & \textbf{Input weights} & \textbf{Indicator relevance}\\
\midrule
\% Over 80 & 0.08 & 0.07\\
$\Delta$ population & 0.08 & 0.04\\
Natural Balance & 0.08 & 0.07\\
\% Over65 living alone & 0.08 & 0.09\\
\% Under18 foreigners & 0.08 & 0.06\\
\% Under 18 - Single parent & 0.08 & 0.07\\
\% Unused dwellings & 0.08 & 0.04\\
\% Graduated & 0.08 & 0.09\\
\% Pop. circulation & 0.08 & 0.08\\
\% people under poverty line & 0.08 & 0.08\\
\% rents & 0.08 & 0.06\\
Median Income (individual) & 0.08 & 0.07\\
Median Income (family) & 0.08 & 0.06\\
\bottomrule
\end{tabular}
    \caption{Average Input and Indicator relevance for the calculation of the autosynth index for Florence, normed dataset.}
    \label{tab:weights_florence}
\end{table}

Table \ref{tab:weights_florence} reports the values for input weights and indicator relevance after the index calculation. All elementary indicators contribute to the SEDI calculation, with two slight exceptions: the share of unused dwellings (\%) and the population change ($\Delta$), both of which appear slightly less relevant than the other variables. We stress that in this work we remain agnostic towards the choice of input weights, thus their value correspond to $\frac{1}{p}$. Moreover, from indicator relevance, calculated according to \ref{eq:posterior} we can notice that the variables that represents more heavily the latent phenomena are the share of graduated people, less prone to social vulnerability, probably with higher revenues and a better social security network, and the share of elderly living alone. However, there are no dramatic differences across the indicator relevances. 

\begin{table}[]
    \centering
\begin{tabular}{lrrrlrrr}
\toprule
\multicolumn{4}{c}{Most vulnerable areas} & \multicolumn{4}{c}{Least vulnerable areas} \\
 & AMPI & AutoSynth & PCA &  & AMPI & AutoSynth & PCA\\
\midrule
Aeroporto & 87.80 & 114.12 & 60.00 & Calatafimi & 97.01 & 60.00 & 102.00\\
S. Jacopino & 100.01 & 114.94 & 119.97 & Bagnese - Fiume Greve & 97.24 & 71.83 & 100.75\\
Peretola & 108.76 & 115.78 & 119.67 & Libertà - Fortezza & 96.15 & 75.33 & 92.60\\
Novoli - Lippi & 106.84 & 116.46 & 117.07 & Cure & 99.38 & 77.03 & 96.02\\
Mercato Centrale & 117.70 & 123.29 & 122.96 & S. Gervasio & 90.38 & 79.32 & 80.43\\
Coverciano & 105.05 & 130.00 & 124.76 & Torre del Gallo & 92.45 & 79.97 & 75.23\\
\bottomrule
\end{tabular}
    \caption{Most and least vulnerable areas in Florence, ranked according the autosynth index}
    \label{tab:top5_florence}
\end{table}

\FloatBarrier

\section {AutoSynth Index in a different Context: Assessing community vulnerability in the U.S.}
%{An example: Assessing community fragility in US}
\begin{comment}
To validate the performance of the AutoSynth index beyond the specific case study that motivated this work, this section presents its application in a different empirical context.
The difference concerns not only the object of the analysis — that is, the specific multivariate phenomenon to be synthesised — but also the characteristics of the data, including a much larger sample size.

Using the dataset provided by \cite{kirkegaard2016inequality} we apply the AutoSynth index to depict community fragility across the continental U.S. counties.

In this section, we illustrate our proposal for synthetic index construction using the procedure described in Section~\ref{sec:method} on a real dataset, that depicts the community fragility among the US  continental counties. We use this example to compare the results obtained with those from two other methods for constructing synthetic indices:  the AMPI \citep{mazziotta2018measuring}, and a composite index based on PCA \citep{mishra2007comparative}.

Community fragility is a cross-border phenomenon, implying that conditions contributing to fragility are not confined solely to individual counties. Rather, fragility in one area often exerts spillover effects on neighbouring regions, generating a multiplicative dynamic that amplifies regional vulnerability.
\end{comment}

This section applies AutoSynth to a second empirical context to assess its performance beyond the case study that motivated this work. While the Florence application is the \emph{motivating case study}, where the substantive framework is developed following the Italian sub-municipal vulnerability literature, the U.S.\ application serves as a \emph{methodological validation}: its purpose is to test AutoSynth on a dataset with fundamentally different characteristics --- 
N=3,136 counties versus 
N=74 suburban units, a different data structure, and independently defined indicators. Accordingly, the dimensional structure of the Community Vulnerability Index (CVI), encompassing economic, social, health, and cultural domains reflects the structure of the available dataset.
Community vulnerability is a cross-border phenomenon: conditions in one county often spill over into neighbouring regions, generating dynamics that amplify vulnerability at the regional scale.
%This interconnectedness can result in geographic pockets of heightened vulnerability, making it crucial to identify these patterns visually and analytically.
Our goal is to identify areas within the US exhibiting the highest levels of vulnerability and to assess the regions experiencing the greatest levels of socioeconomic deprivation nationwide. Such analysis can also offer valuable insights to support more targeted and effective policy interventions. Also in this case, the results are evaluated by comparing them with those obtained using the two reference indices: the AMPI \citep{mazziotta2018measuring} and a composite index based on PCA \citep{mishra2007comparative}.

We examined four main domains representing specific dimensions for CVI: economic, social, health, and cultural vulnerability. These dimensions collectively represent the broader phenomenon of vulnerability and frequently overlap. Specifically, the correlation between community vulnerability and health conditions is particularly relevant and well-documented in the literature \citep{cutter2003social, khazanchi2020county}. Communities experiencing heightened vulnerability often show poorer health outcomes due to limited access to healthcare resources \citep{bourgois2017structural, gaynor2020social}, higher exposure to environmental hazards \citep{fekete2009validation, lehnert2020}, and increased prevalence of chronic diseases \citep{yu2020social}. Understanding these correlations is critical, as it highlights the importance of integrated interventions aimed at simultaneously addressing vulnerability and improving health outcomes.

The presence of correlated indicators poses a significant challenge for researchers attempting to construct synthetic indices. Omitting highly correlated variables might reduce redundancy but simultaneously risks losing important information. In such cases, methods that effectively capture overall variance, such as those based on dimensionality reduction, allow researchers to retain comprehensive information without forcing drastic compromises in the dataset.

\begin{table}[]
    \centering
\begin{tabular}{ccrrrrr}
\toprule
\textbf{Domain} & \textbf{Elementary Index} & \textbf{Mean} & \textbf{st.dev} & \textbf{5\%} & \textbf{Median} & \textbf{95\%}\\
\midrule
Cultural & \% No High School & 16.89 & 7.34 & 7.40 & 15.40 & 30.40\\
& \% Graduate & 6.44 & 3.85 & 2.70 & 5.30 & 13.90\\
& \% Sch. Enroll & 74.97 & 5.06 & 66.60 & 75.15 & 82.75\\
\hline
Economic & Earnings - 2010\$  & 25448 & 5062 & 19002 & 24813 & 34872\\
 & \% Poverty - All & 15.46 & 6.37 & 6.90 & 14.65 & 27.21\\
 & Gini Index & 0.43 & 0.04 & 0.37 & 0.42 & 0.49\\
 & \% Unemployment & 0.07 & 0.02 & 0.03 & 0.07 & 0.12\\
\hline
Social & \%White & 78.81 & 19.60 & 38.48 & 86.35 & 97.40\\
 & \%Afro-american & 8.78 & 14.40 & 0.10 & 1.95 & 41.41\\
 & \% Poverty - 65+ & 11.48 & 5.47 & 5.15 & 10.25 & 22.01\\
 & \% Poverty - 6- & 24.85 & 11.87 & 7.40 & 23.77 & 47.12\\
 & \% Child - Single parent & 31.62 & 9.90 & 16.43 & 30.60 & 49.25\\
\hline
Health & Obesity rate & 0.31 & 0.04 & 0.23 & 0.30 & 0.37\\
 & Uninsured rate & 0.18\textbf{} & 0.05 & 0.10 & 0.18 & 0.27\\
\bottomrule
\end{tabular}
    \caption{Descriptive statistics for elementary indicators employed}
    \label{tab:des_usa}
\end{table}

Drawing on the dataset provided by \cite{cdcsvi2019}, we rely on information for 3,136 U.S. counties, updated to 2019, encompassing 14 distinct dimensions. 
Table \ref{tab:des_usa} presents selected descriptive statistics for these dimensions.
Economic vulnerability is proxied by the median earnings in the county, the Gini coefficient as a measure of unequal distribution of wealth, the unemployment rate, and the overall population living under the poverty threshold. Cultural vulnerability is measured by the share of school enrollment, the share of graduate degrees, and the share of people who left school before completing high school. Moreover, health vulnerability is represented by the share of obesity and the share of uninsured inhabitants. Finally, social vulnerability is captured through the ethnic composition, and the share of children and the elderly living in poverty. Table \ref{tab:weights_usa} reports the input weights and the indicator relevance, calculated accondingly equation \ref{eq:posterior}. As we pointed out, we opt for uniform weights for all the elementary indicators, so their weights corresponds to $\frac{1}{p}$, with $p$ the number of elementary indicators. From ex-post indicator relevance instead we can see that the \% of uninsured people was the most salient to construct the CVI, probably grasping one of the most alarm bell for social vulnerability: the absence of a large share of population that cannot afford medical care. 

\begin{table}[]
    \centering
\begin{tabular}{lrr}
\toprule
\textbf{Elementary Index}  & \textbf{Input weights} & \textbf{Indicator relevance}\\
\midrule
\% No High School & 0.07 & 0.08\\
\% Graduate & 0.07 & 0.06\\
\% Sch. Enroll & 0.07 & 0.08\\
Earnings - 2010\$ & 0.07 & 0.05\\
\% Poverty - All & 0.07 & 0.05\\
Gini Index & 0.07 & 0.05\\
\% Unemployment & 0.07 & 0.05\\
\% White & 0.07 & 0.09\\
\% Afro-american & 0.07 & 0.07\\
\% Poverty - 65+ & 0.07 & 0.05\\
\% Poverty - 6- & 0.07 & 0.06\\
\% Child - Single parent & 0.07 & 0.05\\
Obesity rate & 0.07 & 0.06\\
Uninsured rate & 0.07 & 0.18\\
\bottomrule
\end{tabular}
    \caption{Average Input and Indicator relevance for the calculation of the autosynth index for US counties, normed dataset.}
    \label{tab:weights_usa}
\end{table}

\begin{table}[]
    \centering
    \resizebox{\textwidth}{!}{
\begin{tabular}{lrrrlrrr}
\toprule
\multicolumn{4}{c}{Most vulnerable areas} & \multicolumn{4}{c}{Least vulnerable areas} \\
 & AMPI & AutoSynth & PCA &  & AMPI & AutoSynth & PCA\\
\midrule
Jefferson County, Mississippi & 108.25 & 121.50 & 123.34 & Los Alamos County, New Mexico & 78.15 & 70.00 & 71.29\\
Allendale County, South Carolina & 108.61 & 121.67 & 124.27 & Falls Church city, Virginia & 80.96 & 72.69 & 70.00\\
Humphreys County, Mississippi & 108.85 & 121.92 & 127.15 & Loudoun County, Virginia & 81.13 & 74.41 & 78.24\\
Wilcox County, Alabama & 108.52 & 122.11 & 126.19 & Douglas County, Colorado & 81.48 & 75.54 & 79.50\\
Holmes County, Mississippi & 109.60 & 123.48 & 128.55 & Fairfax County, Virginia & 82.27 & 76.00 & 80.65\\
East Carroll Parish, Louisiana & 113.56 & 130.00 & 129.43 & Arlington County, Virginia & 82.81 & 76.07 & 79.17\\
\bottomrule
\end{tabular}
}
    \caption{Most and least vulnerable counties, ranked according to the autosynth index}
    \label{tab:top5_usa}
\end{table}

\begin{figure}
    \centering
    \includegraphics[width=.7\linewidth]{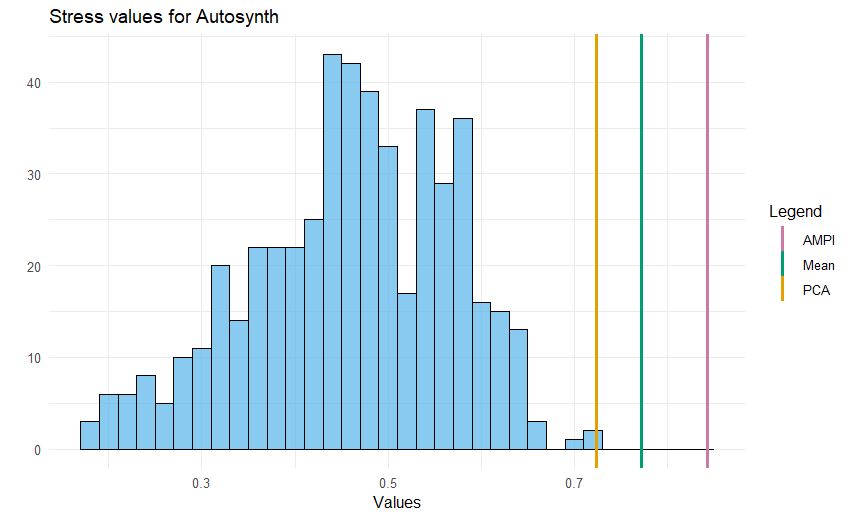}
    \caption{Stress values for autosynth index, compared to the other methods - normed data}
    \label{fig:stress_norm_usa}
\end{figure}

 \begin{figure}
     \centering
     \includegraphics[width=1\textwidth]{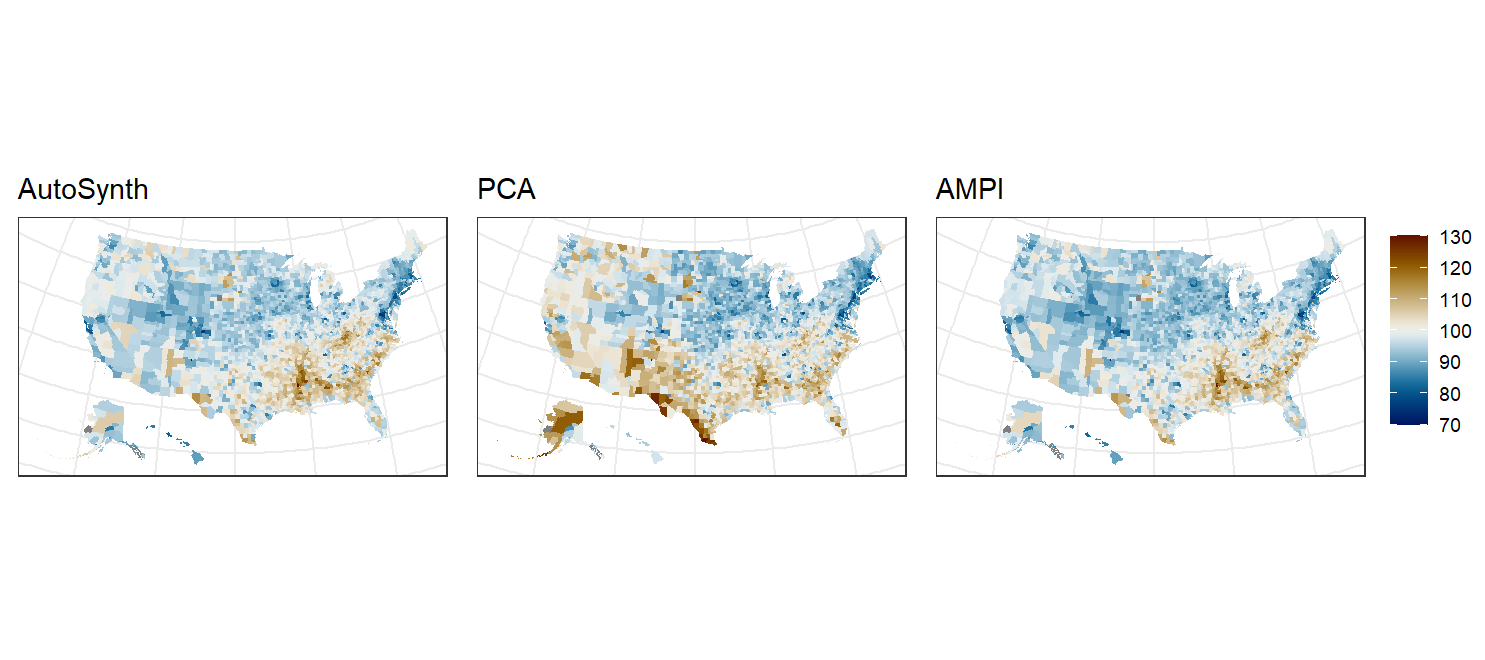}
     \caption{AMPI, PCA and AutoSynth vulnerability Index for US counties, normed data}
    % \label{fig:usmap}

        \centering
     \includegraphics[width=\textwidth]{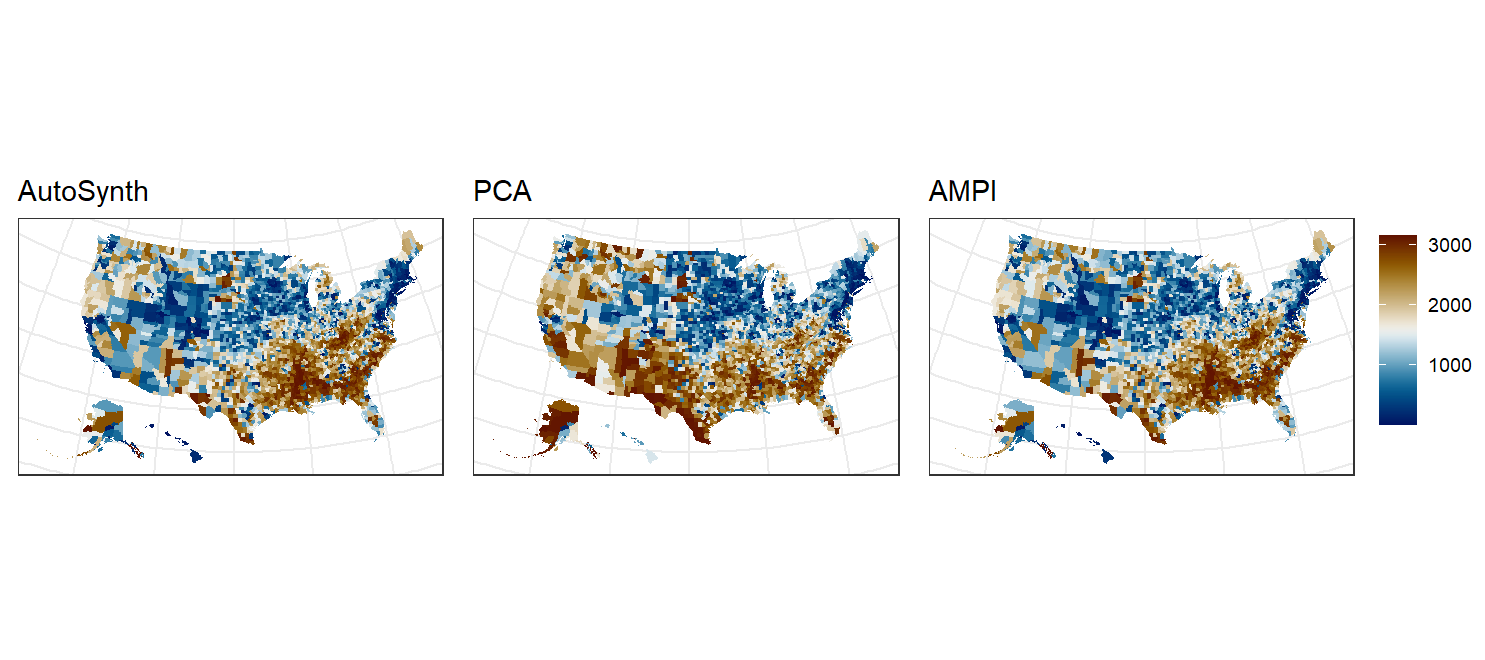}
     \caption{Rank statistics for the AMPI, PCA and AutoSynth vulnerability Index for US counties, normed data}
     \label{fig:usmap}
 \end{figure}

Results are obtained by iteratively applying the AutoSynth procedure to the original dataset, scaled according to equation \ref{eq:norm}. We repeat the estimation process 500 times, here we report the median value for the index distribution.
Figure \ref{fig:usmap} (on the left)  reports the  Community vulnerability index calculated with AutoSynth at the county level in the US for 2019, in absolute value and ranks. 
Notably, the index reveals a clear spatial pattern and identifies distinct clusters of vulnerability, specifically: 

\begin{itemize}
    \item The urban area of New York, Washington and Philadelphia seems to show the lowest levels of vulnerability, as well as the New England area and the Boston area. We can expect this result, as these are the most developed areas of the US. 
    \item On the other hand, the south bend of the US spanning from the Carolinas to Texas exhibits a higher level of vulnerability, especially in Mississippi and Louisiana. 
    \item Midwest, Central US and Rocky Mountain states show lower levels of vulnerability.
\end{itemize}

Table \ref{tab:top5_usa} reports the six most fragile counties and the six least fragile counties in the US. Remarkably, half of the most vulnerable ones are in Mississippi, while the other three lie in neighbouring states, confirming the cross-border hypothesis. 
On the other hand, two-thirds of the least vulnerable counties are in Virginia, highlighting the existence of a low vulnerability area in the Atlantic coast and New England.
These results sound comparable with our expectations and with the results from \cite{svidataset}. We find that the wealthier areas are also the ones that exhibit lower levels of vulnerability, while more depressed counties seem to suffer from multiple sources of vulnerability. 

Turning now to the evaluation of the AutoSynth methodology by benchmarking it against alternative aggregation approaches used to construct synthetic indices, the comparison of the resulting Community vulnerability Index across different methods reveals remarkably consistent geographic patterns, with areas of highest vulnerability remaining stable regardless of the aggregation technique applied. This finding provides strong evidence of the robustness and reliability of the proposed method compared to existing alternatives in the literature.
In terms of stress values, as depicted in Figure \ref{fig:stress_norm_usa}, Autosynth consistently exhibits lower stress values compared to its competitors, producing an improved representation of the euclidean distance between the representation of elementary indicators from US counties. 
To illustrate, a median  stress value of \texttt{0.45} for AutoSynth, 
compared with \texttt{0.72} for PCA and \texttt{0.83} for AMPI 
(Figure \ref{fig:stress_norm_usa}), indicates that the one-dimensional representation 
produced by AutoSynth preserves the pairwise Euclidean distances 
among counties in the original 14-dimensional indicator space more 
faithfully than either alternative. In practical terms, if two 
counties are far apart in their composite socio-economic profiles, 
AutoSynth is more likely to reflect this distance in the final 
index, reducing the distortion introduced by the dimensionality 
reduction.
This further confirms the effectiveness and robustness of the proposed method.
A comparison between the application of AutoSynth in this case and its implementation in the Florence study discussed in Section \ref{sec:florence} can yield additional insights into the performance and adaptability of the AutoSynth index.  We can notice that the stress performances over the US dataset are lower compared with the stress performances in the Florentine case. This results was expected, as the US counties dataset is forty times larger than the Florentine one, and this richness of observation help the estimation of the index. Thus we can advise to use Autosynth especially in presence of large and complex dataset, even if the performances are remarkable even in presence of few observations.
 % \textcolor{red}{un commento su questi rirultati rispetto a quelli di Firenze?}

 % \textcolor{red}{scriviamo che il campione è più grande e abbiamo risultati migliori rispetto a firenze}
\FloatBarrier

\section{Simulations}\label{sec:sims}
In this section, we present a simulation study designed to evaluate the information compression capacity of the AutoSynth procedure across a range of scenarios, varying in the characteristics and interrelations of the underlying elementary indices. The performance of AutoSynth is benchmarked against three alternative synthetic aggregation methods: the arithmetic mean, the AMPI, and a composite index based on PCA. Please refer to section \ref{sec:overview} for a review of these aggregation methods.  

In both applications to real data considered in the previous sections, we observed that the main advantage of the AutoSynth Index suggested in this paper, compared to AMPI and the PCA-based index, is its capability to represent the input elementary indicators, thereby better reproducing the original dimensions within a single feature space. To investigate this property more rigorously, we designed a simulation study spanning scenarios in which the elementary indices vary both in their distributional characteristics and in the strength and form of their interrelationships. Specifically, $3$ distinct data-generating processes (DGPs) were considered in this study, and for each, $3$ different sample sizes were used.
More in detail, the three DGPs, all including fourteen variables representing the elementary indices, are as follows:

\begin{itemize}
    \item \textbf{IID variables}: The elementary indicators are independent and identically distributed, with a Normal distribution 
    \item \textbf{Correlated ID variables}: The elementary indicators are correlated, but all follow the same distribution, which, as in the previous case, is the normal distribution. The correlations among the elementary indicators range from -0.87 to 0.79.
    \item \textbf{Correlated, no ID distributions}: The elementary indicators are correlated and follow different distributions: two uniform distributions, a $\chi^2$
distribution, a Poisson distribution, an exponential distribution, a Student's t-distribution, and three normal distributions. The correlations among the elementary indicators range between -0.85 and 0.91.
\end{itemize}
The three values chosen for the sample sizes are 50, 250, and 1000. For each of the 3x3 combinations of DGP and sample size, the number of replications is 1000.
The hyperparameters of the distributions have been simulated from appropriate uniform distributions.
We opt for pre-treating the elementary indicators by scaling them with equation \ref{eq:norm}. In this way, we are not altering the shape of the distribution, but are rescaling the units in a common range to avoid the variability of an elementary indicator prevailing over the others and affecting the final representation of the latent phenomena.

% \textcolor{red}{normalizzazione?} 
\begin{figure}
    \centering
    \includegraphics[width=0.9\linewidth]{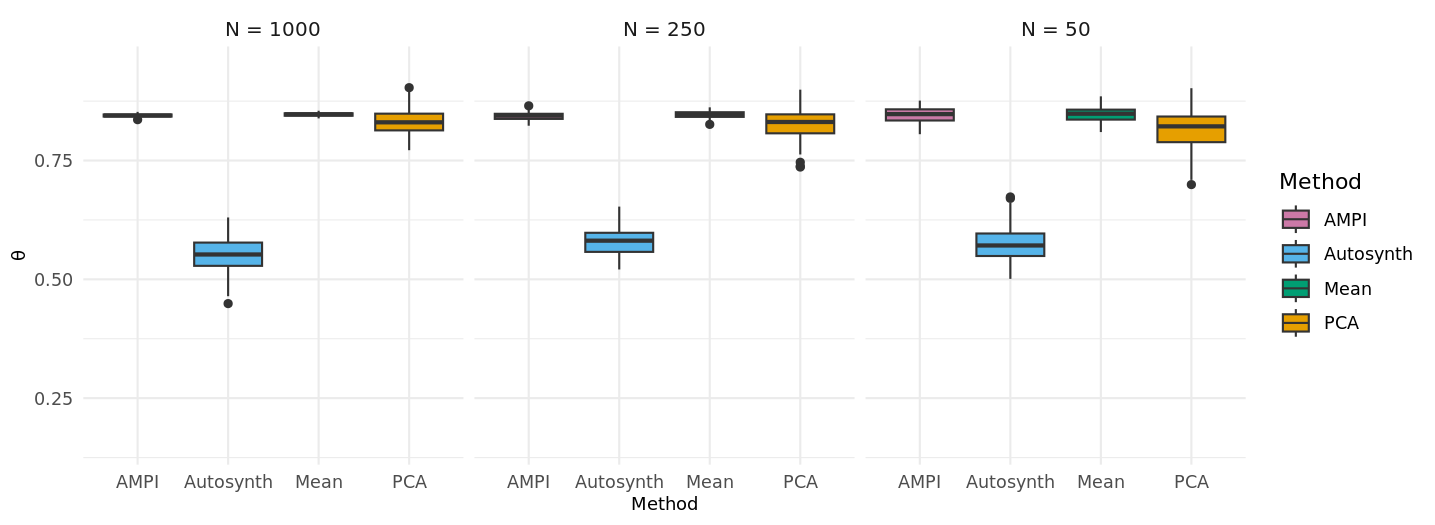}
    \caption{Stress values for the synthetic index - IID case.}
    \label{fig:1_dgp}
\end{figure}

\begin{figure}
    \centering
    \includegraphics[width=0.9\linewidth]{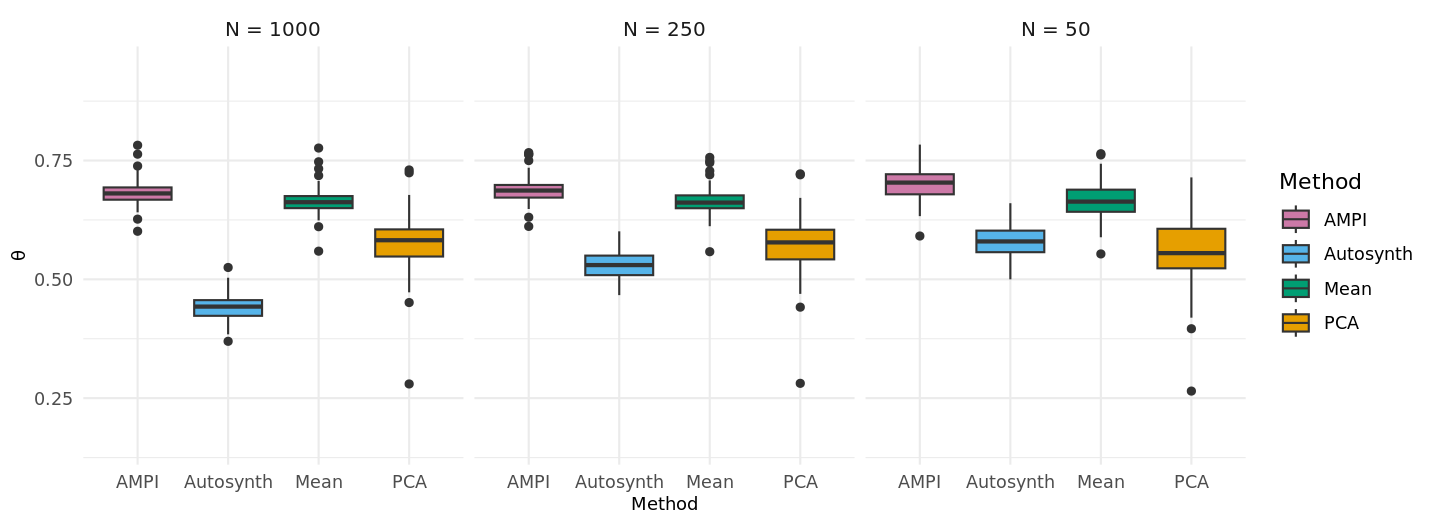}
    \caption{Stress values for the synthetic index - Non independent, ID case.}
    \label{fig:2_dgp}
\end{figure}

\begin{figure}
    \centering
    \includegraphics[width=0.9\linewidth]{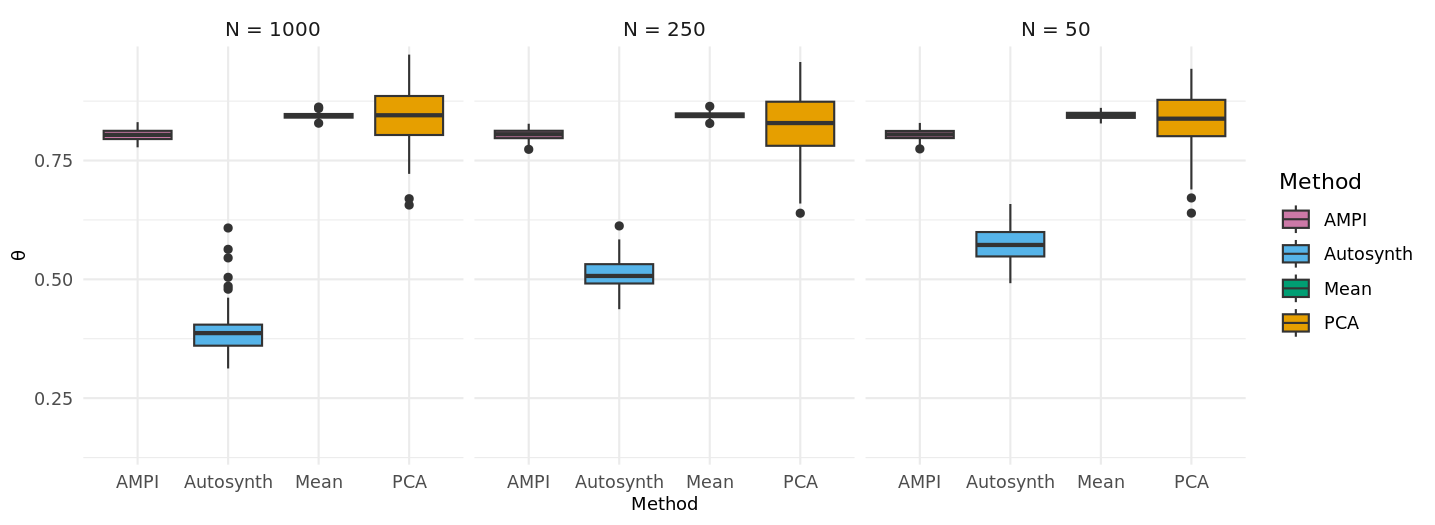}
    \caption{Stress values for the synthetic index - Non independent nor ID case.}
    \label{fig:3_dgp}
\end{figure}

In all simulation scenarios, AutoSynth achieves lower median stress values than competing methods in preserving inter-observational distances. However, the magnitude of the improvement varies with sample size: for small samples ($n = 50$) the gain over conventional methods depends on the dependency structure and independence across variables (see Figure \ref{fig:2_dgp} and Figure \ref{fig:3_dgp}). This gain becomes increasingly pronounced as $n$ grows, a trend not observed in competing algorithms. Moreover, being based on neural network optimisation, AutoSynth is inherently stochastic: repeated runs on the same data produce a distribution 
of solutions rather than a unique index. While this variability is 
mitigated by iterating the procedure and reporting a summary 
statistic (we use the median of 500 runs), estimation instability 
is more pronounced with small samples. When the performance 
advantage over simpler, deterministic methods is marginal and the 
resulting rankings are comparable, the additional complexity may not 
be warranted. 
Accordingly, while AutoSynth is competitive
regardless of sample size, its adoption is especially advantageous 
for large and complex datasets, where its flexibility yields 
substantial and stable gains.

\begin{comment}
    In all simulation scenarios, the results reinforce the insights from our empirical analyses. Under stress‐test conditions, AutoSynth consistently outperforms benchmark methods in preserving inter‐observational distances. Moreover, its performance advantage grows as sample size increases—a trend not seen in competing algorithms. Accordingly, we recommend relying on conventional aggregation techniques for small‐sample studies and adopting AutoSynth as the dataset size expands. 
\end{comment}

Finally, as shown in Figure \ref{fig:3_dgp}, AutoSynth excels at aggregating variables with heterogeneous distributions across a variety of data configurations.
 In the Appendix, we report in figures \ref{fig:1_dgp_r}, \ref{fig:2_dgp_r}, \ref{fig:3_dgp_r}, the variability of the stress values applied to the ranks instead of being applied to the dataset, in our idea this measure represents the \textit{stability} of the estimation procedures.

\section{Conclusions}

In this study, we introduced AutoSynth, an innovative methodology for the construction of composite indices, leveraging the capabilities of autoencoders. This approach distinguishes itself through its data-driven dimensionality reduction, effectively addressing the limitations inherent in traditional linear methods such as PCA. Through a series of rigorous simulation studies and applications to real-world datasets, we have demonstrated AutoSynth's capacity to accurately capture complex, non-linear relationships within data, thereby providing a more precise and meaningful representation of multidimensional phenomena.

The flexibility of AutoSynth is particularly evident in its ability to manage heterogeneous data, characterised by varying distributions and diverse sample sizes. This characteristic renders it exceptionally suitable for the analysis of intricate socio-economic phenomena, where variables often exhibit disparate behavioral patterns. Furthermore, the provision for incorporating expert-defined input weights facilitates the integration of prior knowledge into the index construction process, enhancing the relevance and accuracy of the resulting composite indices.

Moreover, this work introduces a method for constructing the elementary indicators relevance as the reconstruction error in the synthetic index. This value allow us to understand which indicators are more salient to represent the latent phenomenon, fostering its understanding. 
We leverage Autosynth to depict the vulnerability of communities into two main examples: the SEDI index for calculating vulnerability in Florentine suburbs and the CVI index to assess vulnerability across US counties.

The application of our proposed AutoSynth methodology to assess socio-economic and demographic vulnerability in the sub-municipal areas of Florence revealed distinct patterns of vulnerability across the city. Notably, the historic city center and western Florence emerged as areas with the highest levels of vulnerability, a finding consistent with previous studies. This counterintuitive result for the historic center can be attributed to gentrification and the proliferation of short-term rentals, which have altered the area's socio-economic landscape. Conversely, the eastern and southern suburbs, known for their scenic beauty, exhibited the lowest levels of vulnerability. Comparative analysis with traditional methods, such as AMPI and PCA, demonstrated that AutoSynth produced comparable results, particularly in ranking the areas, while exhibiting lower stress values. This suggests that the ordering of sub-municipal areas by vulnerability is robust across different aggregation methods, with AutoSynth providing a more accurate representation of the underlying data structure. The identified patterns of vulnerability underscore the complex interplay of socio-economic and demographic factors in Florence, highlighting the need for targeted policy interventions to address these disparities.

AutoSynth analysis of U.S. county vulnerability revealed distinct patterns: urban areas showed low vulnerability, while the southern belt exhibited high vulnerability, particularly in Mississippi and Louisiana. AutoSynth outperformed traditional methods in representing inter-county distances, confirming its effectiveness in assessing community vulnerability. 

Finally, we study the empirical properties of the proposed aggregation method with a simulation study in which we test several different DGPs and sample sizes, stressing how AutoSynth is a non-inferior choice to common aggregation methods, which outperforms the three alternative methods (arithmetic mean, AMPI, PCA) when the sample size is large or when the elementary indicators' distribution is non-i.i.d..

Despite the promising outcomes, it is imperative to acknowledge the limitations of our study. Specifically, further exploration is warranted to optimize the parameter selection of the autoencoder and to evaluate the impact of diverse distance metrics on the results. Additionally, the application of sequential autoencoders for the construction of hierarchical indices represents a promising avenue for future research, potentially enabling the analysis of complex phenomena at varying levels of granularity.

Regarding future perspectives, we posit that AutoSynth possesses significant potential for application across a broad spectrum of domains. Its capacity to synthesise complex information into meaningful indices can be particularly instrumental in informing policy decisions, monitoring progress towards sustainable development goals, and evaluating the impact of multifaceted interventions. Moreover, the integration of AutoSynth with other advanced data analysis techniques, such as predictive modeling and interactive visualization, may unlock new frontiers in the comprehension of multidimensional phenomena.

In conclusion, the AutoSynth methodology represents a substantial advancement in the construction of composite indices, offering a data-driven, flexible, and interpretable approach. Its ability to capture non-linear relationships, manage heterogeneous data, and integrate expert knowledge renders it a valuable tool for the analysis of complex phenomena in diverse contexts.

\noindent\textbf{Acknowledgements} The authors are thankful to Gianni Dugheri and Fiammetta Menchetti for the insightful comments. 

\noindent\textbf{Funding sources} The authors thanks the UNIFI4FUTURE - SPARKLING Grant B17G24000250006. 

\noindent \textbf{Competing interest}: The authors declare that they have no known competing financial interests or personal relationships that could have appeared to influence the work reported in this paper.

\noindent \textbf{Data availability}: Data about this work are either publicly available or available upon request. 

\bibliography{bibtex}% common bib file
\clearpage
\begin{appendices}
\section*{Appendix}
In the appendix we report additional material to the main text. Firstly, we report the schematized of the calculation procedure for AutoSynth, described formally in the main text. Secondly, we provide additional results for the simulation study described in section \ref{sec:sims}. We estimate the stress test described in equation \ref{eq:stress}, but applied to the values of observations' ranks, as following 

\begin{equation}\label{eq:rank_ stress}
 \Theta_R = \sqrt{\frac{\sum_{i=1}^N (r_{ij} - \widetilde{r_{ij}})^2}{\sum_{i=1}^N r_{ij}^2}} \end{equation}

 with $r_{i,j} = \sqrt{\mathcal{R} \frac{1}{p} \sum_{k=1}^p \bm X_i - \mathcal{R} \frac{1}{p} \sum_{k=1}^p \bm X_j)^2}$ and $\widetilde{r_{i,j}}= \sqrt{(\mathcal{R}(\widetilde{\bm Y_i}) - \mathcal{R}( \widetilde{\bm Y_j}))^2}$, where $\mathcal{R}$ is the rank operator.  

\begin{algorithm}
\caption{\textbf{Index Construction Algorithm}}
\begin{algorithmic}
\Require Dataset of elementary indicators $\bX$
\Ensure Constructed composite index $\widetilde{\bY}$.

\State \textbf{Step 1: Variable Selection}\\
According to expert knowledge, main variables should be selected to represent the concept represented in the composite index

\State \textbf{Step 2: Normalization}\\
Elementary indexes should be rescaled to a common range, either via minmax rescaling or through standardization, different normalization choices implies slightly different results.

\State \textbf{Step 3: Aggregation}\\
During aggregation phase the autoencoder is trained to represent the input data. Thus, the estimated encoder $\hat{\phi}$ is used to construct the 'coded' version of the dataset, the composite indicator.

\State \textbf{Step 4: Analysis and Post-Estimation Tuning}
\State 4.1. Analyze the results obtained after rescaling the variables.
\State 4.2. Assess the performance of the composite index in capturing the desired concept or idea.
\State 4.3. Evaluate the index's suitability for its intended purpose, such as decision-making or policy analysis.
\State 4.4. Perform post-estimation tuning, if necessary, to improve the index's performance.
\State 4.5. Tuning may involve adjusting the weightings of variables, modifying the normalization or rescaling process, or incorporating additional expert knowledge.
\end{algorithmic}
\end{algorithm}

\begin{figure}
    \centering
    \includegraphics[width=0.9\linewidth]{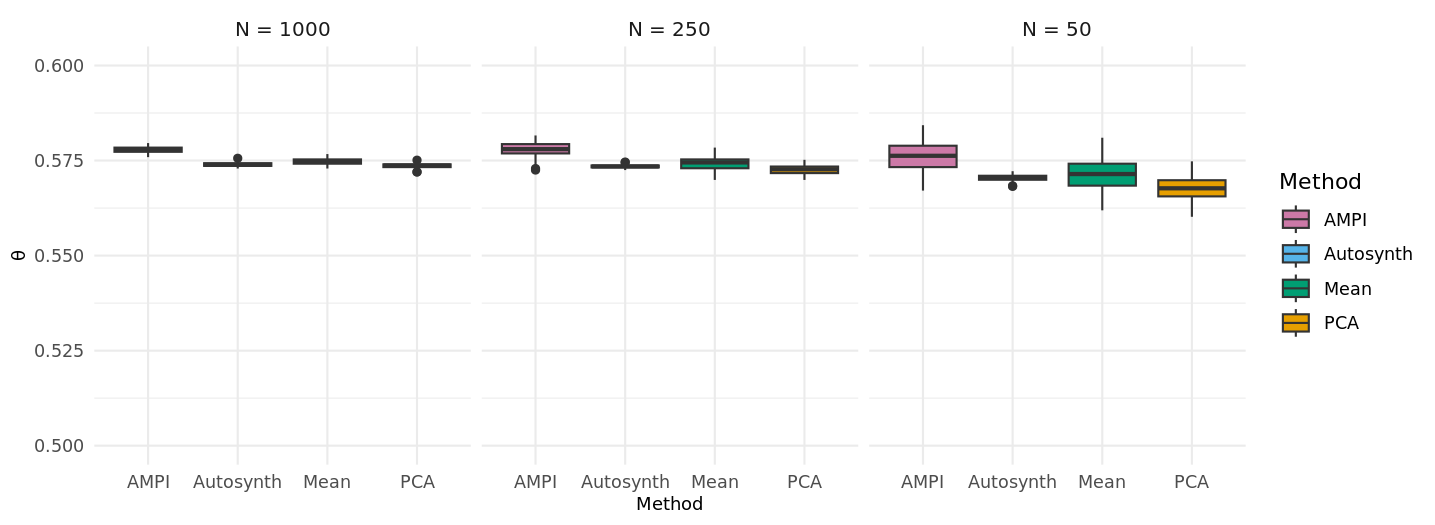}
    \caption{Rank stress values for the synthetic index - IID case.}
    \label{fig:1_dgp_r}
\end{figure}

\begin{figure}
    \centering
    \includegraphics[width=0.9\linewidth]{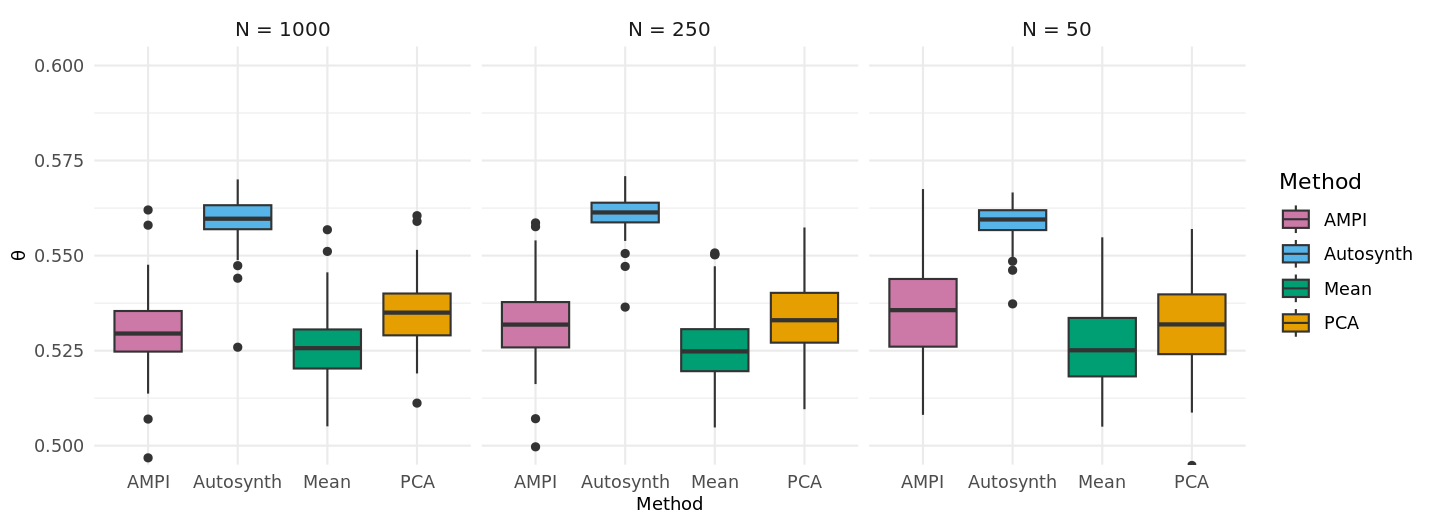}
    \caption{Rank stress values for the synthetic index - Non independent, ID case.}
    \label{fig:2_dgp_r}
\end{figure}

\begin{figure}
    \centering
    \includegraphics[width=0.9\linewidth]{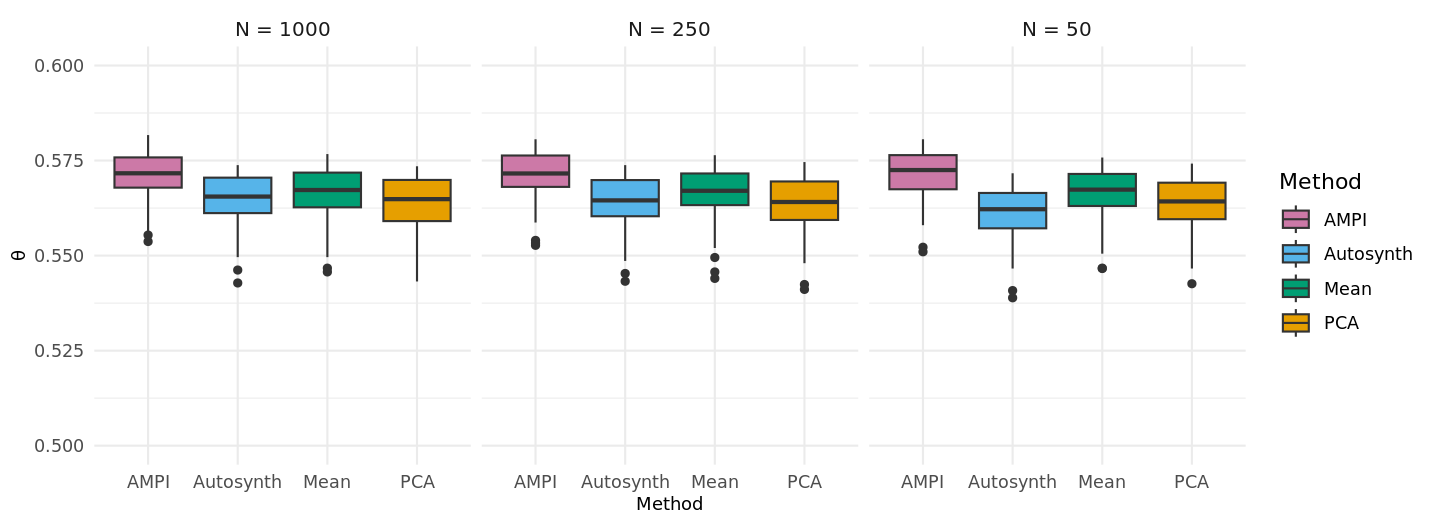}
    \caption{Rank stress values for the synthetic index - Non independent nor ID case.}
    \label{fig:3_dgp_r}
\end{figure} 

\end{appendices}


\end{document}